\documentclass{raa}            % referee version: for submission
\usepackage{graphicx,times}             %for PS/EPS graphics inclusion, new
\usepackage{natbib}
\usepackage{amssymb,amsmath}
\bibpunct{(}{)}{;}{a}{}{,}

\usepackage[pagebackref=true]{hyperref}

\begin{document}

  \title{A geometric distortion solution specifically for historical observations and its implementation
}
%   \subtitle{I. Place Your Subtitle Here}

   \volnopage{Vol.0 (20xx) No.0, 000--000}      %%preserved for Editor. DOn't remove!
   \setcounter{page}{1}          %%starting page, preserved for Editor. DOn't remove!

   \author{F. R. Lin %(������) %% Put your Chinese name in "( )" if you like. Note to open line 11 "\usepackage[UTF8]{ctex}"
      \inst{1,2}
   \and Q. Y. Peng
      \inst{2,4}
   \and Z. J. Zheng
      \inst{2,3}
\and B. F. Guo
      \inst{2,4}
   }


   \institute{School of Software, Jiangxi Normal University, Nanchang 330022, China; {\it frlin@jxnu.edu.cn}\\
        \and
Sino-French Joint Laboratory for Astrometry, Dynamics and Space Science, Jinan University, Guangzhou 510632, China\\
        \and
            Guangdong Ocean University, Zhanjiang 524000, China\\
\and
 Department of Computer Science, Jinan University, Guangzhou 510632, China\\
\vs\no
   {\small Received 20xx month day; accepted 20xx month day}}

\abstract{Geometric distortion (GD) critically constrains the precision of astrometry. Using well-established methods to correct GD requires calibration observations, which can only be obtained using a special dithering strategy during the observation period. Unfortunately, this special observation mode is not often used, especially for the historical observations before those GD correction methods presented. As a result, some telescopes have no GD calibration observations for a long period, making it impossible to accurately determine the GD effect. This limits the value of the telescope observations in certain astrometric scenarios, such as using historical observations of moving targets in the solar system to improve their orbits. We investigated a method for handling GD that does not rely on the calibration observations. With this advantage, it can be used to solve the GD models of telescopes which were intractable in the past. The method was implemented in Python and released on GitHub. It was then applied to solve GD in the observations taken with the 1-m and 2.4-m telescopes at Yunnan Observatory. The resulting GD models were compared with those obtained using well-established methods to demonstrate the accuracy. Furthermore, the method was applied in the reduction of observations for two targets, the moon of Jupiter (Himalia) and the binary GSC2038-0293, to show its effectiveness. After GD correction, the astrometric results for both targets show improvements. Notably, the mean residual between observed and computed position ($O-C$) for the binary GSC2038-0293 decreased from 36 mas to 5 mas.
}

   \authorrunning{Lin et al.}            %author_head in even pages
   \titlerunning{A GD solution for historical observations}  % title_head in odd pages

   \maketitle
%% The author head (on even pages) and the title head (on odd pages) will be
%% automatically extracted from \author{} and \title{}. Whenever the title is too long,
%% you will be asked to supply a shorter one by inserting either \authorrunning{} or
%% \titlerunning{} before \maketitle. Anyway, you can specify your own heads.
%%
%%
%% Note: In the following text body of your manuscript, please note several differences from
%%       other major journals:
%% (1) \subsection{Please Capitalize the First Letter of Each Notional Word in Subsection Title}
%% (2) Please Capitalize the First Letter of Each Notional Word in all tables' captions

%
%________________________________________________ sections below
%

\section{Introduction}
\label{intro}

Both space and ground-based observations in the field of optical astrometry are inevitably affected by geometric distortion (GD). In the majority of scenarios,  the correction of GD is the main factor limiting the astrometric precision \citep{Peng2017,Wang2017,McKay2018,Wang2019,Casetti2021}. For example, for the observations taken with the 2.4-m telescope at Yunnan Observatory, the astrometric precision nearly improved by a factor of 2 after accurate correction of GD \citep{Peng2017}.   

Besides its effect on astrometric precision, GD correction is also necessary for some applications to improve computational speed or meet fundamental requirements. For instance, in real-time applications that involve handling substantial data and computational loads for detecting and tracking near-Earth objects, \citet{Zhai2018} employed GD correction when converting the pixel coordinate to the equatorial coordinate. After GD correction, the iterative process of mapping from the pixel coordinate to the standard coordinate can achieve convergence faster. As a result, this mapping process was accelerated while attaining high-precision results. 

Adopting high-order plate constants for reduction can mitigate the effect of GD. However, it requires sufficient reference stars exist in the field of view (FOV), which is not met in specific high-precision applications. For some ground-based observations of moving targets, instrument limitations and other factors may result in a lack of observed reference stars available from the observations. Consequently, the GD solution derived from calibration observations is essential for obtaining high-precision positions of these moving targets. In addition, in HST extremely deep field observations where the AB magnitude of stars can reach up to 30 mag \citep{Illingworth2013}, there is no star catalogue that provides accurate positions for reference stars. The infeasibility of applying the high-order plate constants, as a result, makes GD correction necessary.

The most notable research on addressing GD was conducted by \citet{Anderson2003}. Their method does not rely on star catalogues and is called the self-calibration technique. It has been applied to solve GD for multiple cameras on the HST \citep{Bellini2009} as well as several ground-based telescopes \citep{Anderson2006,Bellini2010}, achieving high-precision astrometry. An alternative approach for solving GD is based on the known reference star positions that are not affected by GD, such as HST observation positions with GD correction \citep{Service2016}, or even star positions obtained from non-optical wavelength bands \citep{Reid2007}. This method requires fewer observations for GD solution than the self-calibration technique, but the accuracy may be affected by the errors present in the external reference system \citep{Bernard2018}. \citet{Peng2012} proposed a novel method for mitigating the influence of catalogue errors on GD solution, thereby reducing the accuracy requirements of the external reference catalogue. Furthermore, it has been improved by \citet{Wang2019} to handle the GD in observations captured by the 2.3-m Bok telescope at Kitt Peak \citep{Peng2023}.

The above methods can effectively solve the GD model and achieve positional measurements with precision up to the 0.01 pixel level. Due to the GD correction in the reduction, the astrometric precision of natural satellites has been substantially improved \citep{Peng2015,Peng2017,Wang2017}. Nonetheless, these GD correction methods necessitate well-planned observational strategies, obtaining optimally dithered and overlapping frames \citep{Anderson2003,Peng2012}, to offset the effects of GD or catalogue errors \citep{Zheng2021}. Taking these GD calibration observations requires additional telescope time, which is sometimes difficult to meet due to observation conditions. More importantly, for the historical observations before the above methods were proposed, there was no such observation plan to obtain calibration observations at all. That is to say, the above methods cannot deal with the GD in many historical observations.

To address this issue, a method is proposed to derive an analytical GD model without the requirement of GD calibration observations. We demonstrate the performance of this method by comparing it with the well-established GD correction method \citep{Peng2012, Wang2019}. Additionally, we perform the reduction of Himalia (J6) observations captured by the 2.4-m telescope at Yunnan Observatory and the 60-cm telescope observations for a  binary system GSC02038-00293. Himalia is the largest member of Jovian irregular satellites \citep{Grav2015}. We have been observing and measuring the positions of irregular satellites for the past decade \citep{Peng2017,Shang2022}, dedicated to improving astrometric methods to obtain high-precision astrometric results. The high-precision astrometric observations of irregular satellites can be used for improving its ephemerides, as well as understanding the formation of early solar system. GSC2038-00293 is a close binary system with high-level magnetic activity, studying its nature is of great significance for understanding stellar evolution \citep{Dal2012}. Although a lot of research has been done on the system, including light-curve analysis and out-of-eclipse analyses, its nature is still not very clear. A new study intends to combine previous work with the position changes of the binary system to reveal its unknown nature, so we perform astrometry for its historical observations. These observations are initially reduced using plate constants without GD correction, but the results are unsatisfactory. Like many telescopes primarily employed for photometric purposes, the 60-cm telescope at Yunnan Observatory has never performed any GD calibration observations. However, the novel research focus of the binary GSC02038-00293 highlights the importance of astrometry. Since our method does not require GD calibration observations, it is adopted in the reduction of these observations. 

This paper is structured as follows: Section 2 provides detailed information about the observations and the corresponding instruments used to capture them, while Section 3 describes the GD correction method based on the Gaia DR3 \citep{Gaia2023}. Section 4 presents the performance comparison between our method and the well-established method, as well as the advantages of the new method in reducing the observations of the targets J6 and GSC02038-00293. Finally, Section 5 concludes the paper with some closing remarks.

\section{Observations}
\label{sect:obs}
Observations obtained from multiple telescopes are used in this paper. These telescopes include the 60-cm telescope \citep{Zang2022}, the 1-m telescope (IAU code 286, longitude\---E$102^{\circ}47^{\prime}18^{\prime\prime}$, latitude--N$25^{\circ}1^{\prime}46^{\prime\prime}$, and height--2000m above sea level) at Yunnan Observatory, and the 2.4-m telescope (IAU code O44, longitude\---E$100^{\circ}1^{\prime}51^{\prime\prime}$, latitude--N$26^{\circ}42^{\prime}32^{\prime\prime}$, and height--3193m above sea level) at Yunnan Observatory (YNO 60-cm, YNO 1-m, and YNO 2.4-m). More instrumental details of the reflectors and CCD detectors are listed in Table~\ref{tbl1}. The patterns and magnitudes of GD experienced by these instruments are different.

\begin{figure} 
	\center
	\includegraphics[width=  0.4 \columnwidth]{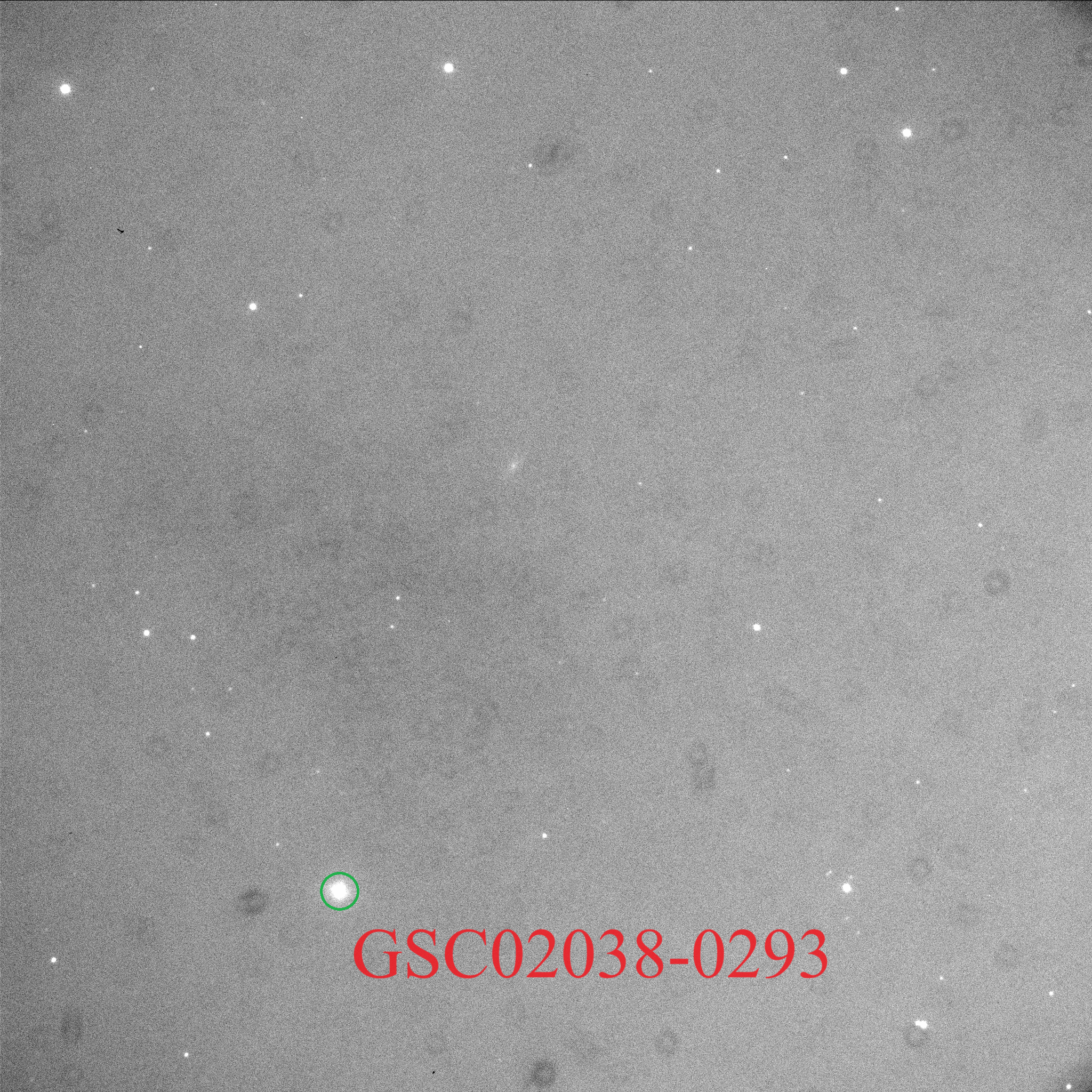}  \\ 
    \caption{A frame of GSC02038-0293 observations obtained from the YNO 60-cm telescope.}
    \label{fig:gsc60cm} 
\end{figure}

Observation sets 1 and 2 were captured using the dithering strategy, which takes multiple dithered exposures of the same sky field with different offsets \citep{Peng2012}. They are used to demonstrate that the proposed method in this paper achieves the same accuracy as other well-established GD correction methods. Observation sets 3 and 4 are significantly affected by higher-order GD, but only a dozen or so bright stars can be used to solve the plate constants, thus the GD solution is very important for high-precision astrometry of the targets. The detailed information of these observations is provided in Table~\ref{tbl2}. In this paper, bright stars refer to stars brighter than a certain magnitude threshold. The astrometric precision of stars brighter than this threshold is consistent, which means their precision is not limited by their magnitude. For our observations, the signal-to-noise ratio (SNR) of the star corresponding to the threshold is about 100, so it can serve as a criterion for using SNR to determine bright stars. This threshold may vary due to some factors such as atmospheric turbulence, so the specific criterion for determining bright stars depend on each observation set. Figure~\ref{fig:gsc60cm} presents a frame of GSC02038-0293 observations obtained from the YNO 60-cm telescope.

\begin{table*}
\begin{center}
\caption{Specifications of the 1-m and 2.4-m telescopes and the corresponding CCD detectors.\label{tbl1}}
\scalebox{1}{
\begin{tabular}{@{}l*{15}{l}}
\hline\hline
Parameter&1-m telescope&2.4-m telescope&60-cm telescope\\
\hline
Approximate focal length&1330 cm &1920 cm &750 cm \\
F-Ratio & 13.1 &8.0 &12.5\\
Diameter of primary mirror&101.6 cm & 240.0 cm &60.0 cm\\
Approximate scale factor & 0.234 arcsec pixel$^{-1}$ & 0.283 arcsec pixel$^{-1}$ & 0.383 arcsec pixel$^{-1}$\\
Size of CCD array (effective)& 4096 $\times$ 4112&1900 $\times$ 1900 &2048 $\times$ 2048 \\
Size of pixel &15.0 $\mu m\ \times$ 15.0 $\mu m$ & 13.5 $ \mu m\ \times$ 13.5 $\mu m$& 13.5 $ \mu m\ \times$ 13.5 $\mu m$\\
\hline
\end{tabular}
}
\end{center}
\end{table*} 

\begin{table*}
\begin{center}
	\caption{Details of the observations\label{tbl2}. The first column is the identification of the observation set. Column (2) lists the target. Column (3) and Column (4) give the right ascension and declination of the target. Column (5) lists the observational date. Column (6) gives the number of CCD frames in each observation set. Column (7) is the range of zenith distance. Column (8) is the telescope used. Column (9) provides the number of stars with sufficient signal-to-noise ratio, whose astrometric precision is not dominated by centering errors. Column (10) gives the exposure time.
\label{tbl2}
}
\scalebox{0.9}{
\begin{tabular}{@{}c*{15}{c}}
\hline\hline
ID&Target &R.A.&Dec.&Obs Date&Frames&Z.D.&Telescope&Bright Stars &Exposure \\
  &        &(degree)&(degree)&(y-m-d) & (No.)  &  (degree)  &  & (No.)  & (second)\\
 (1)&(2)&(3)&(4)&(5)&(6)&(7)&(8)&(9)&(10) \\
\hline
1  & NGC2168      &92.32 &+24.33&  2018-11-13 &38 & 1$\sim$17&YNO 1-m &   350$\sim$400   &60 \\
2  & NGC1664      &72.71 &+43.56&  2015-02-10 &44 & 19$\sim$21&YNO 2.4-m &    150$\sim$180    &40 \\ 
3  & Himalia      &197.02 &-5.15&  2017-04-08 &15& 35$\sim$37&YNO 2.4-m    & $\sim$10    &30 \\
4  & GSC02038-00293 &240.70& +25.34&  2011-02-28 &187&1$\sim$30&YNO 60-cm   & $\sim$12    &40 \\
\hline
\end{tabular}
}
\end{center}
\end{table*}

\section{Methods}
\label{sect:methods}
The method investigated in this paper derives an analytical GD model, which is characterized by a high-order polynomial, using the distortionless positions of stars provided by the Gaia catalogue.
The analytical GD model can effectively describe the GD effect in ground-based observations, because the majority of GD components can be characterized by polynomials. The remaining components, which are typically described using a lookup table \citep{Wang2019}, generally only account for a minor portion of GD. This is confirmed in the subsequent section through experimentation. The principle of this method is to extract the GD effect present in each frame of observations, and then derives the GD model based on the GD effect extracted from these multiple frames. In other words, the method uses the weighted average of the plate constants to derive the GD model. As fitting errors are eliminated by averaging the coefficients from multiple frames, the final GD solution does not have overfitting issues even if only a dozen bright stars can be used to solve the high-order polynomial.

The implementation details of this method are as follows. A two-dimensional Gaussian fitting is used to determine the pixel positions of the observed stars. These observed stars are then cross-matched with the stars given in the Gaia catalogue \citep{Gaia2023} to obtain their reference positions. Specifically, the reference positions are topocentric astrometric positions of the stars \citep{Kaplan1989} calculated from their catalogue positions. To ensure the accuracy of the GD solution, we also account for additional factors that may cause deterioration to its accuracy. These factors include differential colour refraction and charge transfer efficiency issues, which can be effectively addressed using the method presented in \citet{Lin2020}. Consequently, we can obtain the pixel coordinate $(x_i,y_i)$ and the equatorial coordinate $(\alpha_i,\delta_i)$ of each star $i$. The standard coordinate $(\xi_i,\eta_i)$ can be converted from the equatorial coordinate via the central projection, which is presented in \citet{Green1985}.%\citet[][see Equation (1) in that paper]{Peng2012}.

To extract the GD effect on pixel positions, we need to solve a six-parameter linear transformation to obtain the approximate pixel positions $(x_i^{_{L}},y_i^{_{L}})$ of the reference stars.
The linear transformation is expressed as:
\begin{equation}
\begin{gathered}
\label{linear}
x_i = \widetilde{a}\xi_i + \widetilde{b}\eta_i+\widetilde{c}, \\
y_i=  \widetilde{d}\xi_i + \widetilde{e}\eta_i+\widetilde{f},  
\end{gathered}
\end{equation}
where the coefficients $\widetilde{a}$$\sim$$\widetilde{f}$ (denoted as $C^{_{L}}_{\text{std}}$ hereafter) can be estimated through the least-squares fitting. Using the linear transformation, the standard coordinates $(\xi_i,\eta_i)$ can be converted to the approximate pixel positions $(x_i^{_{L}},y_i^{_{L}})$. The coefficients of the linear transformation $C^{_{L}}_{\text{std}}$ are initially inaccurate because they are affected by the GD. During the iterative solving process of GD, the pixel positions $(x_i,y_i)$ in Equation~\ref{linear} will be replaced by the positions after GD correction in each new iteration. As a result, the approximate pixel positions $(x_i^{_{L}},y_i^{_{L}})$ would converge to the distortionless pixel positions.

Based on the pattern and magnitude of GD experienced by the optical system of each telescope, we select a polynomial of appropriate order $N$ to characterize its analytical GD model. The general formula of the polynomial is given as:
\begin{equation}
\begin{gathered}
\label{poly}
U =\sum_{ m,n,0 \leqslant  m+n \leqslant N} k_{mn} X^m Y^n, \\
V=\sum_{ m,n,0 \leqslant  m+n \leqslant N} j_{mn} X^m Y^n,
\end{gathered}
\end{equation}
where $k_{mn}$ and $j_{mn}$ are the parameters to be fitted. Setting $(X,Y)$ as coordinates $(x_i^{_{L}},y_i^{_{L}})$ and $(U,V)$ as coordinates $(x_i,y_i)$, an $N$th-order polynomial that characterizes the GD effect can be fitted. We denote the coefficients of this polynomial as $C_{\text{pix}}$. By solving for the coefficients $C_{\text{pix}}$ of each frame in an observation set and applying a weighted average based on image quality, an average GD solution $\overline{C}_{\text{pix}}$ can be obtained. Most of the random errors are offset in the weighted average of the information from multiple frames, leaving only the GD effect.

\begin{figure*}
\begin{center}

	\includegraphics[width=0.9 \columnwidth]{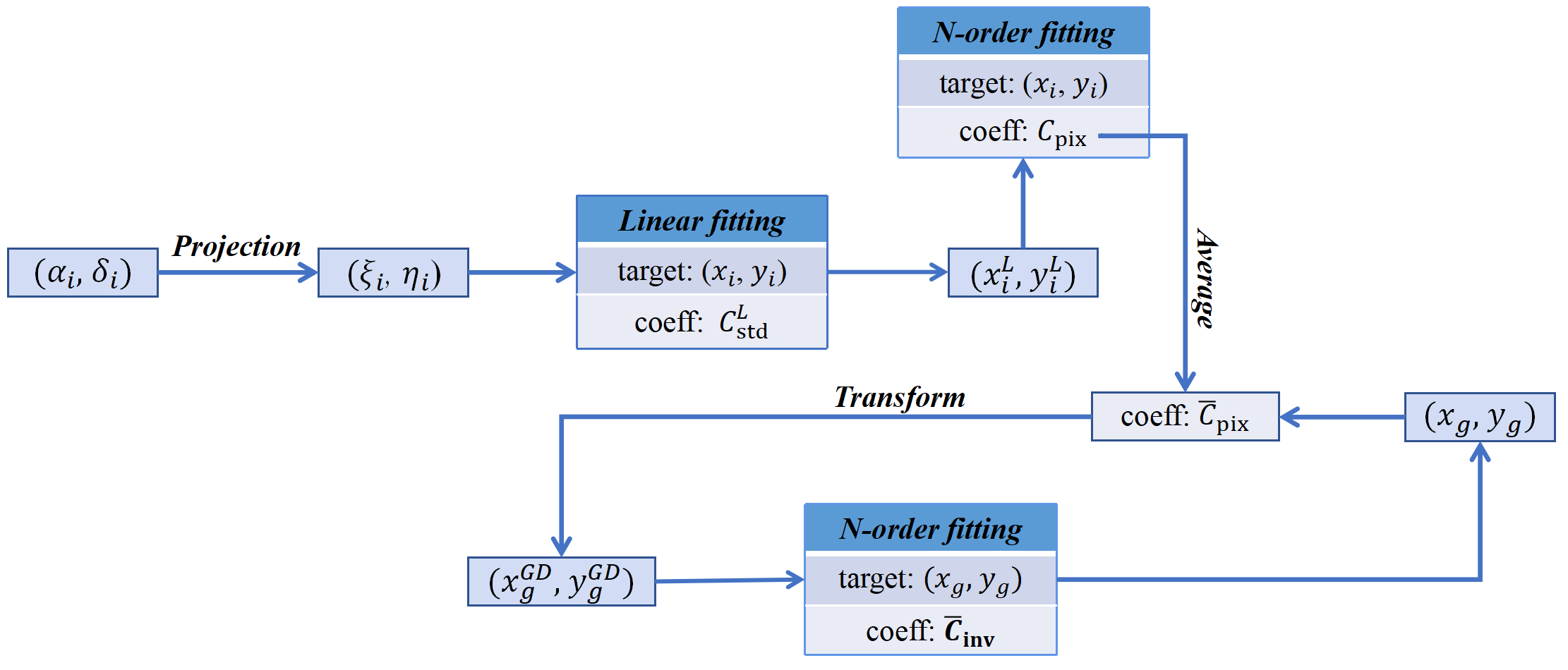} 
    \caption{The transformations between different positions and the solving process of the corresponding coefficients in one iteration of our geometric distortion (GD) solution.  
The arrows in the figure indicate the processes of coordinate transformations through the given coefficients, as well as other operational processes denoted in bold. The \emph{target} in the figure represents the fitting targets of the input positions, while the \emph{coeff} denotes the fitted coefficients obtained from the fitting process. }
    \label{fig:flowchart}
\end{center}
\end{figure*}

Now we can determine the GD effect at any given pixel position using a polynomial with coefficients $\overline{C}_{\text{pix}}$. However, when the GD effect changes dramatically within a small image range, there would be a significant difference in the GD effect between the distortionless pixel position of the star and its actual observed pixel position. In order to handle this issue, we determine the transformation from the pixel positions $(x_i,y_i)$ to the distortionless positions to correct GD. Specifically, we construct a $16\times16$ grid uniformly distributed across the pixel coordinates of the image (as shown in Figure~\ref{fig:gd}). Then the grid positions $(x_g,y_g)$ are transformed via a polynomial using the coefficients $\overline{C}_{\text{pix}}$, resulting in their distorted positions $(x_g^{_{GD}},y_g^{_{GD}})$. Finally, the inverse transformation coefficients $\overline{C}_{\text{inv}}$ are determined by fitting from $(x_g^{_{GD}},y_g^{_{GD}})$ to $(x_g,y_g)$. The pixel position with GD correction can be calculated by setting $(X,Y)$ as the coordinate $(x_i,y_i)$ and using $\overline{C}_{\text{inv}}$ as the coefficients in Equation~\ref{poly}. Figure \ref{fig:flowchart} describes the solving process for these coefficients and the transformations between different positions.

Considering that the pixel positions of stars are contaminated by different levels of random noise, weights have been introduced into all fitting procedures related to the pixel positions $(x_i,y_i)$. The GD model is derived through an iterative procedure, with the weights initially set to be uniform. After the first iteration of the GD solution, the weight for each star is determined as the inverse of the variance in positional measurements.
The variance can be calculated by fitting a sigmoidal function to the Mag-SD data (as shown in Figure~\ref{fig:precision}). The sigmoidal function is expressed as:
\begin{equation}
\sigma(m)=(A_1-A_2)/(1+e^{(m-m_0)/dm})+A_2,
\end{equation}
where $m$ is the magnitude of the star, $\sigma(m)$ is the positional SD of the star in the previous iteration, $A_1 ,A_2, m_0$ and $dm$ are the free parameters. Detailed calculation procedure for the weights can be found in \citet{Lin2019}. The weights and the coefficients $C^{_{L}}_{\text{std}}$ are updated in each iteration, so that a more accurate GD model can be solved. The final analytical GD model is obtained through two to four iterations of the aforementioned procedure. For the convenience of others to use the method to solve GD, a Python implementation of this method is available on GitHub\footnote{\url{https://github.com/JxnuLin/GDSolver}}.

For comparison, a classical method of the plate constants reduction is also applied in this paper. The method can be simply described as solving a polynomial transformation from pixel positions of reference stars to their standard positions, and then using the transformation to calculate the astrometric position of the target. Using a $N$th-order polynomial for reduction, GD not higher than $N$th-order can be handled if there are enough reference stars \citep{Green1985,Peng2010}.

%For example, the 3rd-order plate constants can be obtained by fitting the coefficients of Equation~\ref{poly} when $N$ takes 3. 

\section{Results}
Compared with the GD model determined by the well-established method \citep{Peng2012, Wang2019}, the accuracy of the GD model obtained through our method is verified. Furthermore, our method is applied to reduce observations of Himalia (J6) and GSC02038-00293 to demonstrate its advantages. The computed positions of J6 and GSC02038-00293 are retrieved from Institute de M$\rm\acute{e}$chanique C$\rm\acute{e}$leste et de Calcul des $\rm\acute{E}$ph$\rm\acute{e}$m$\rm\acute{e}$rides (IMCCE) and Gaia DR3, respectively.

\subsection{Comparison with the well-established method}

\begin{figure*}
\begin{center}
	\includegraphics[width=1  \columnwidth]{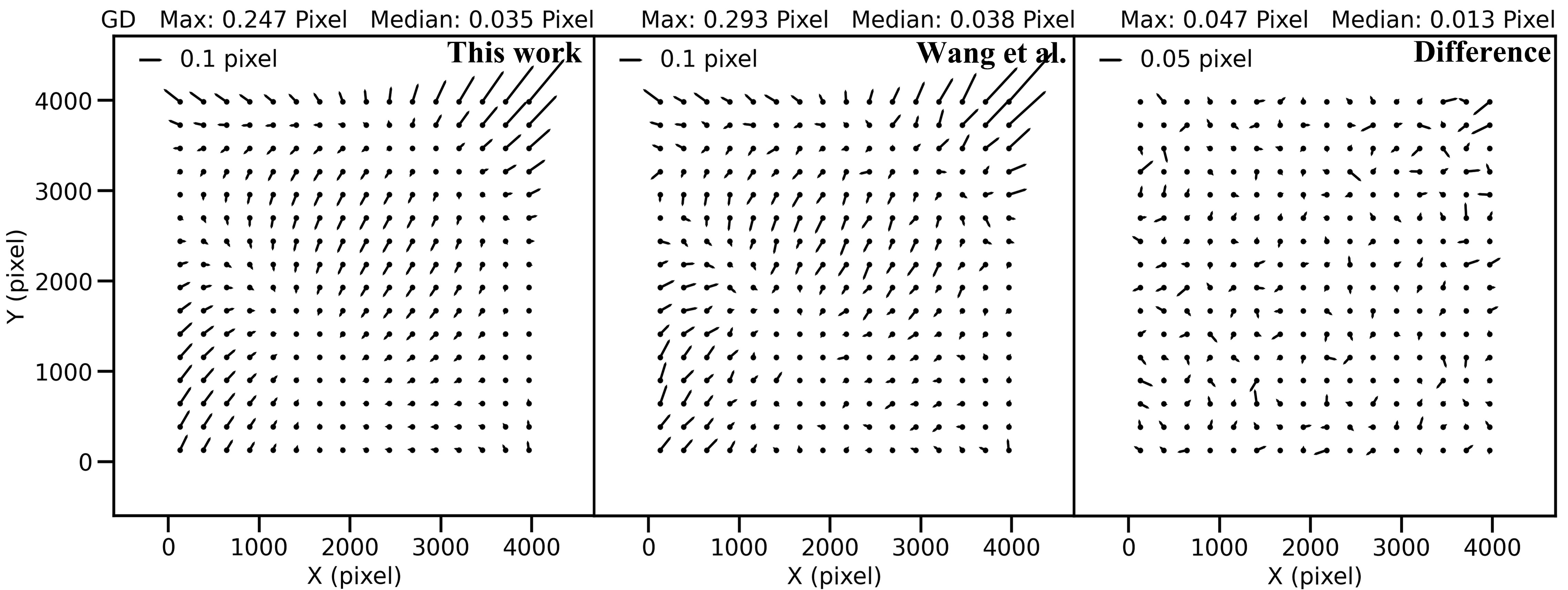} 
	~\\ \includegraphics[width=1   \columnwidth]{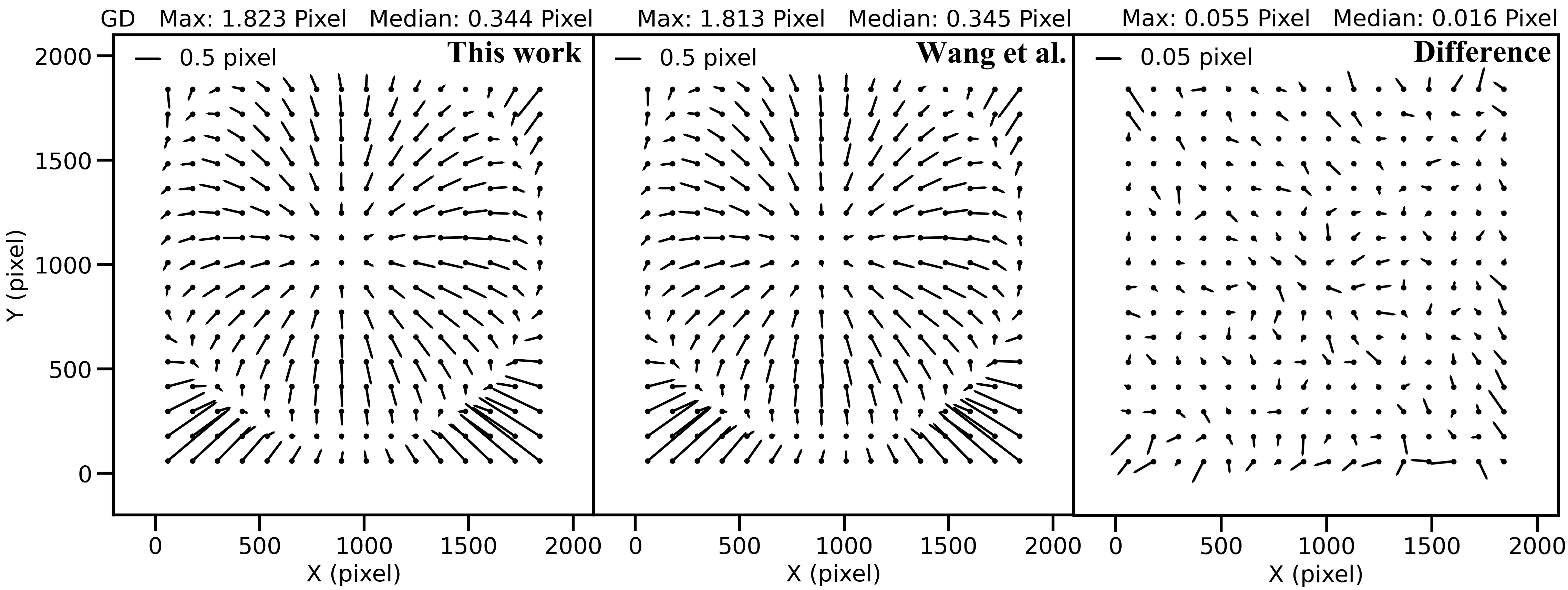}  
    \caption{The GD models of the YNO 1-m and 2.4-m telescopes solved by our method and the method of \citet{Wang2019}, respectively. The differences of these two GD solutions for each telescope are presented in the right panels. The upper panels display the GD models of the YNO 1-m telescope, while the lower panels show the GD models of the YNO 2.4-m telescope. The statistics of the GD vectors are presented at the top of each panel. The vectors are suitably magnified to clearly visualize the shape of the GD.}
    \label{fig:gd}
\end{center}
\end{figure*}

Since observation sets 1 and 2 were acquired by the dithering strategy, the well-established methods can also be used to solve GD. The GD models for these observation sets were solved by the methods described in Section~\ref{sect:methods} and \citet{Wang2019}, respectively. Figure \ref{fig:gd} presents the results, which include the GD models for the YNO 1-m and 2.4-m telescopes solved by each method. The differences of the GD models solved by these two methods for each telescope are also given in the right panels of the figure. Among them, the analytical GD model for the YNO 1-m telescope is characterized using a 4th-order polynomial, while the model for the YNO 2.4-m telescope using a 5th-order polynomial. 

\begin{figure*}
\begin{center}
	\includegraphics[width=0.48 \columnwidth]{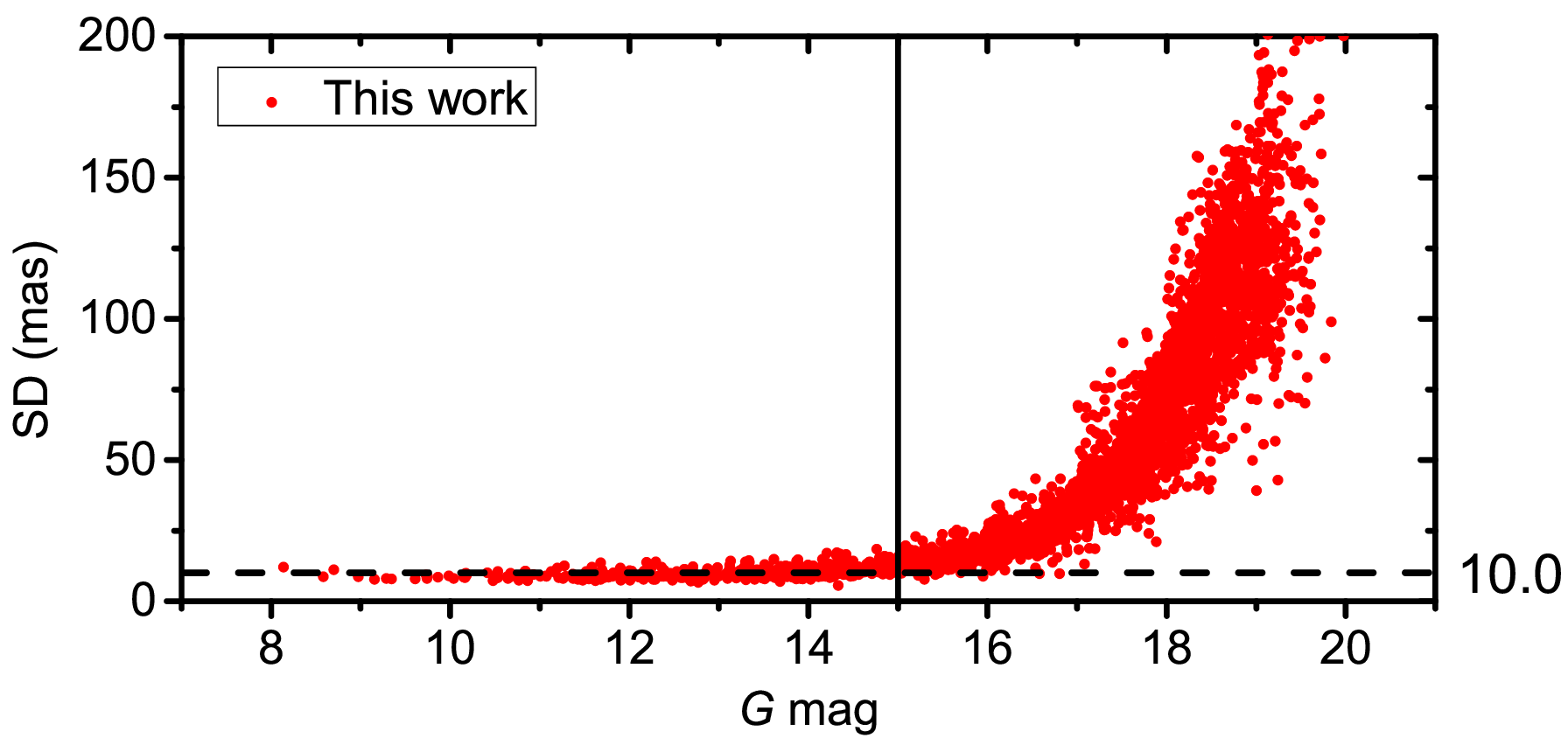} \quad 
	\includegraphics[width=0.48 \columnwidth]{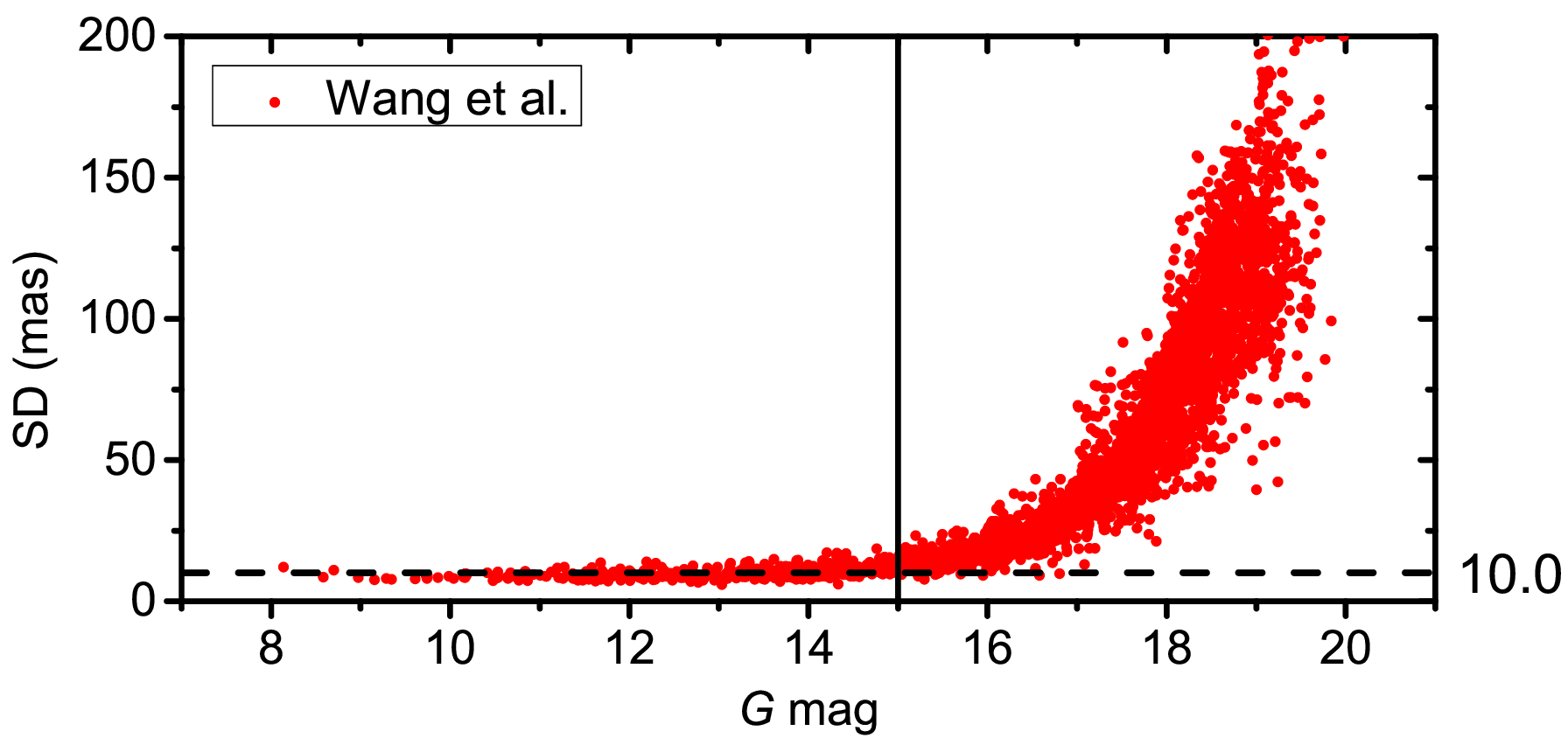}      ~\\~\\ 
	\includegraphics[width=0.48 \columnwidth]{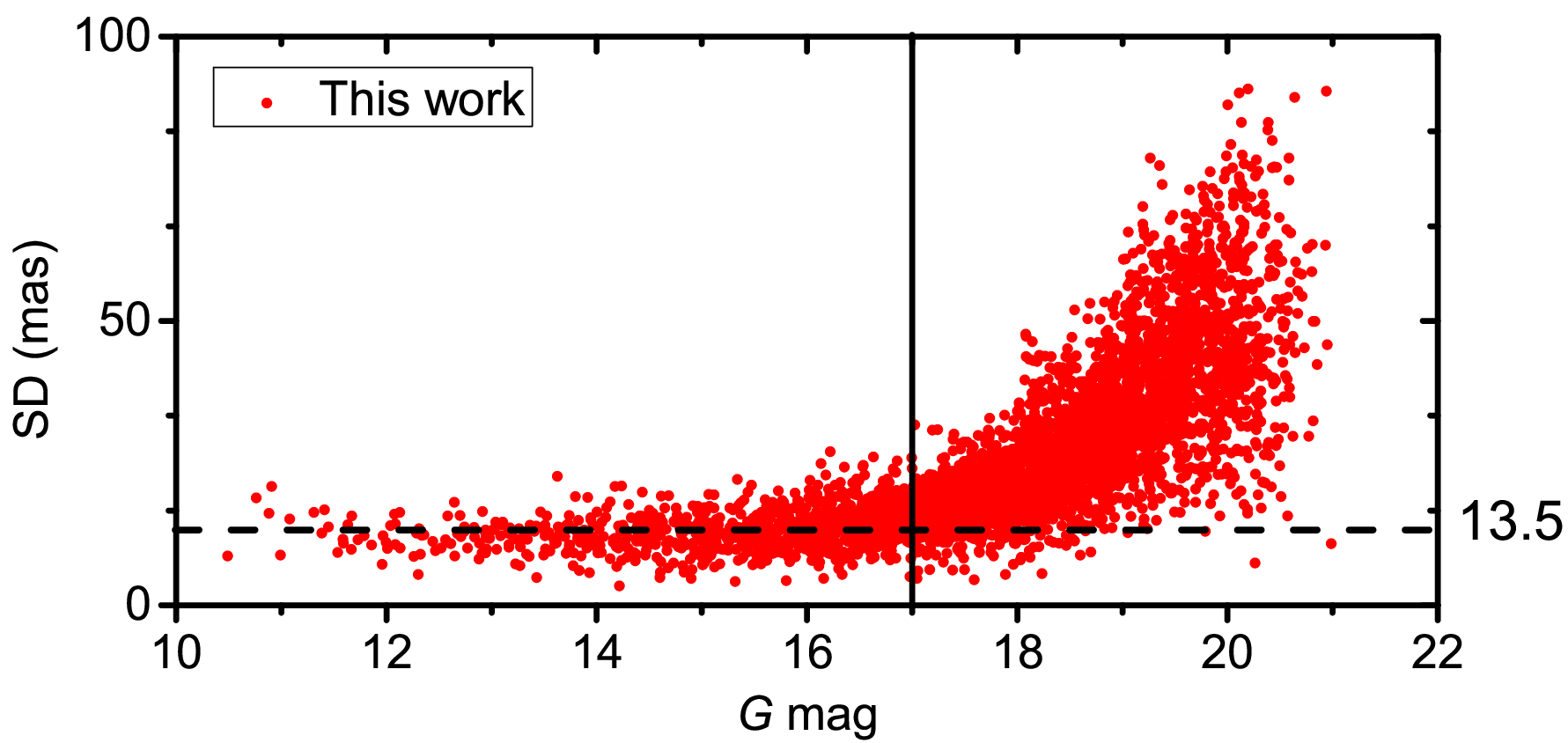}  \quad
	\includegraphics[width=0.48 \columnwidth]{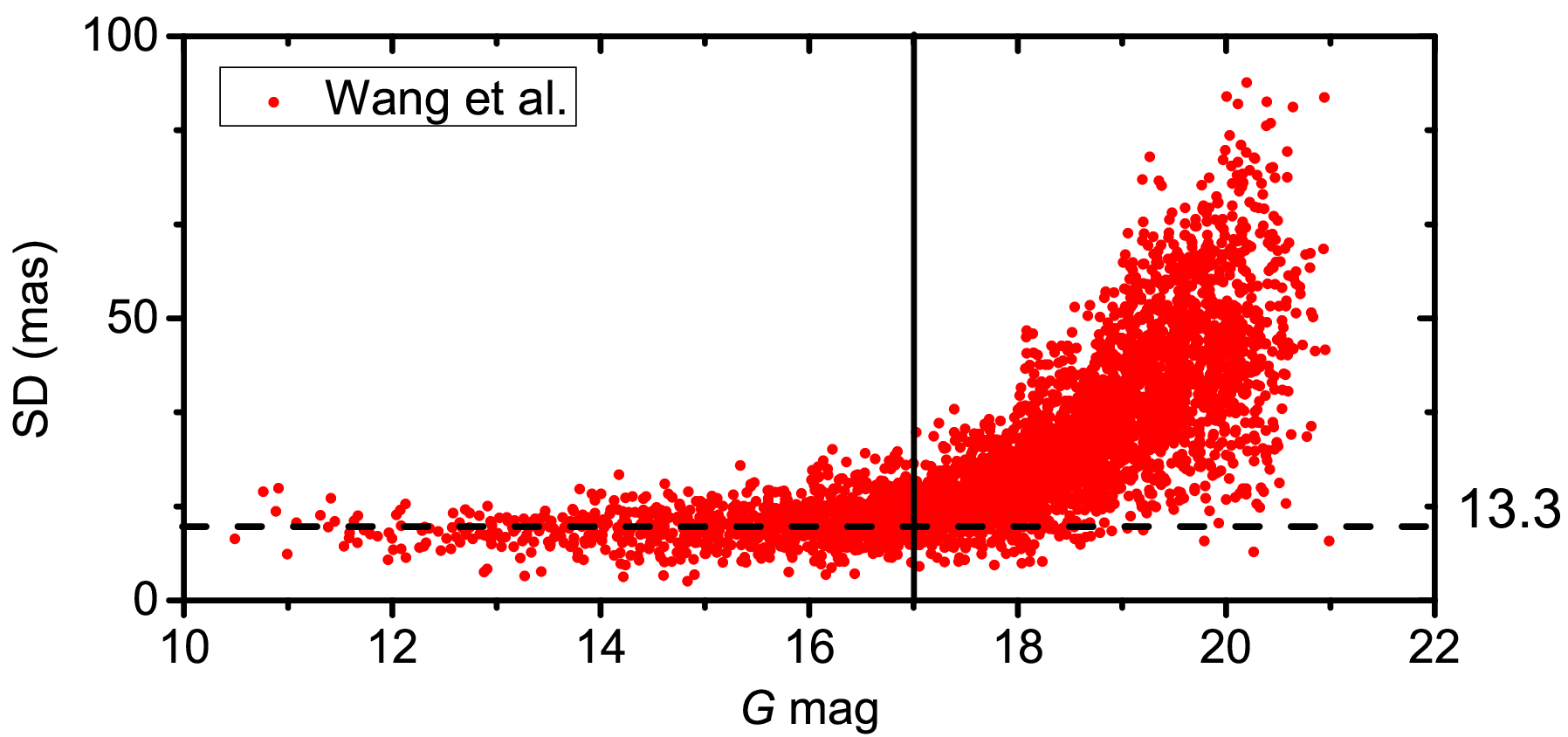}      
    \caption{Comparison of the positional standard deviation (SD) obtained after GD correction using our method versus the reference method \citep{Wang2019}. The upper panels present the results for observation set 1 captured by the YNO 1-m telescope, and the bottom panels show the results for observation set 2 captured by the YNO 2.4-m telescope. The horizontal axis is Gaia G-mag and the vertical axis is the positional SD calculated by $\sigma_{\text{sum}}=\scriptstyle{\sqrt{\sigma_{\alpha \cos \delta}^2+\sigma_\delta^2}}$. The horizontal dashed line marks the median of the positional SDs for stars brighter than the magnitude indicated by the vertical line, the number on the right is the median.}
    \label{fig:precision}
\end{center}
\end{figure*}

\begin{figure*}
\begin{center}
	\includegraphics[width=0.47  \columnwidth]{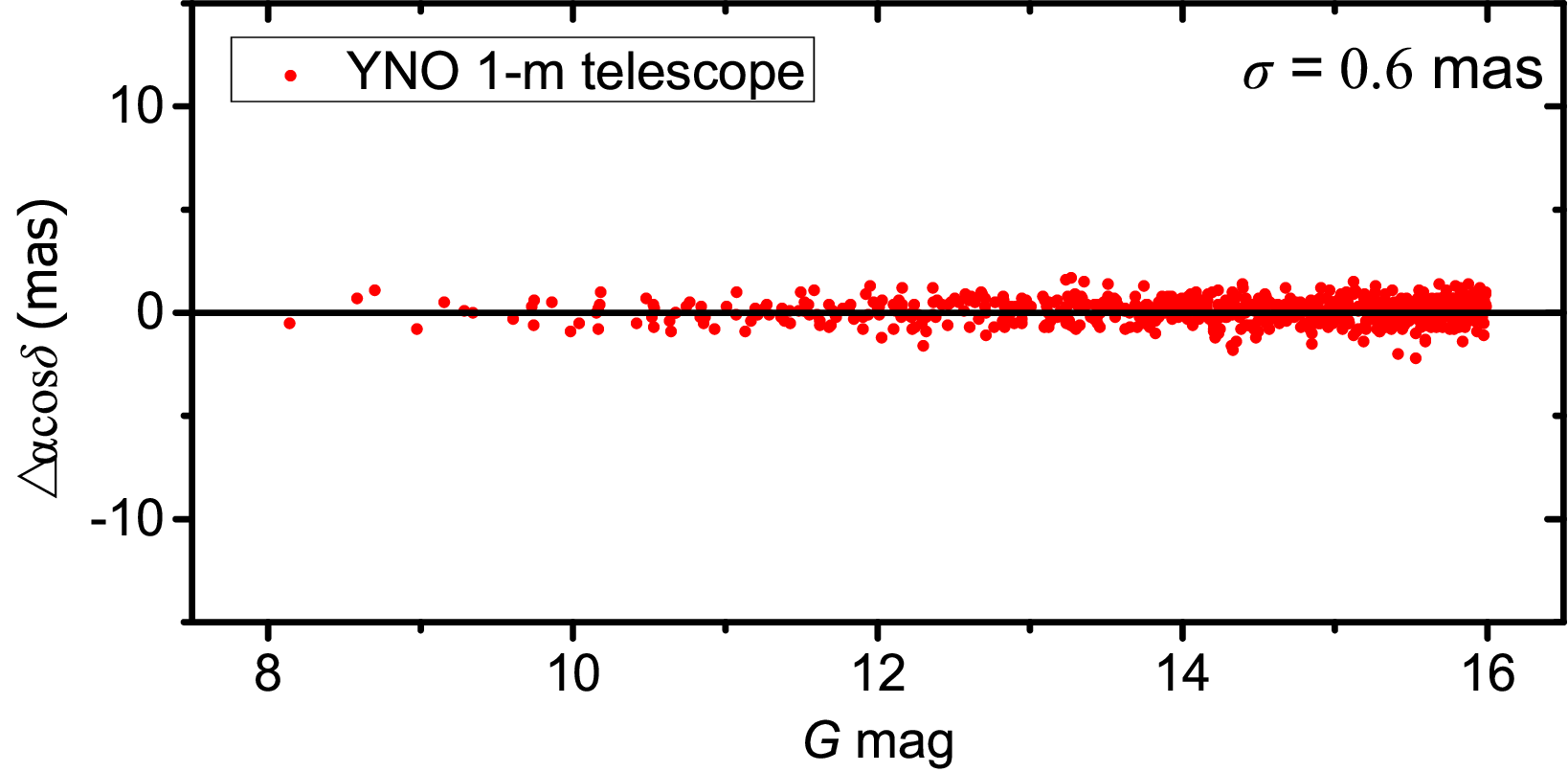} 
	\includegraphics[width=0.47  \columnwidth]{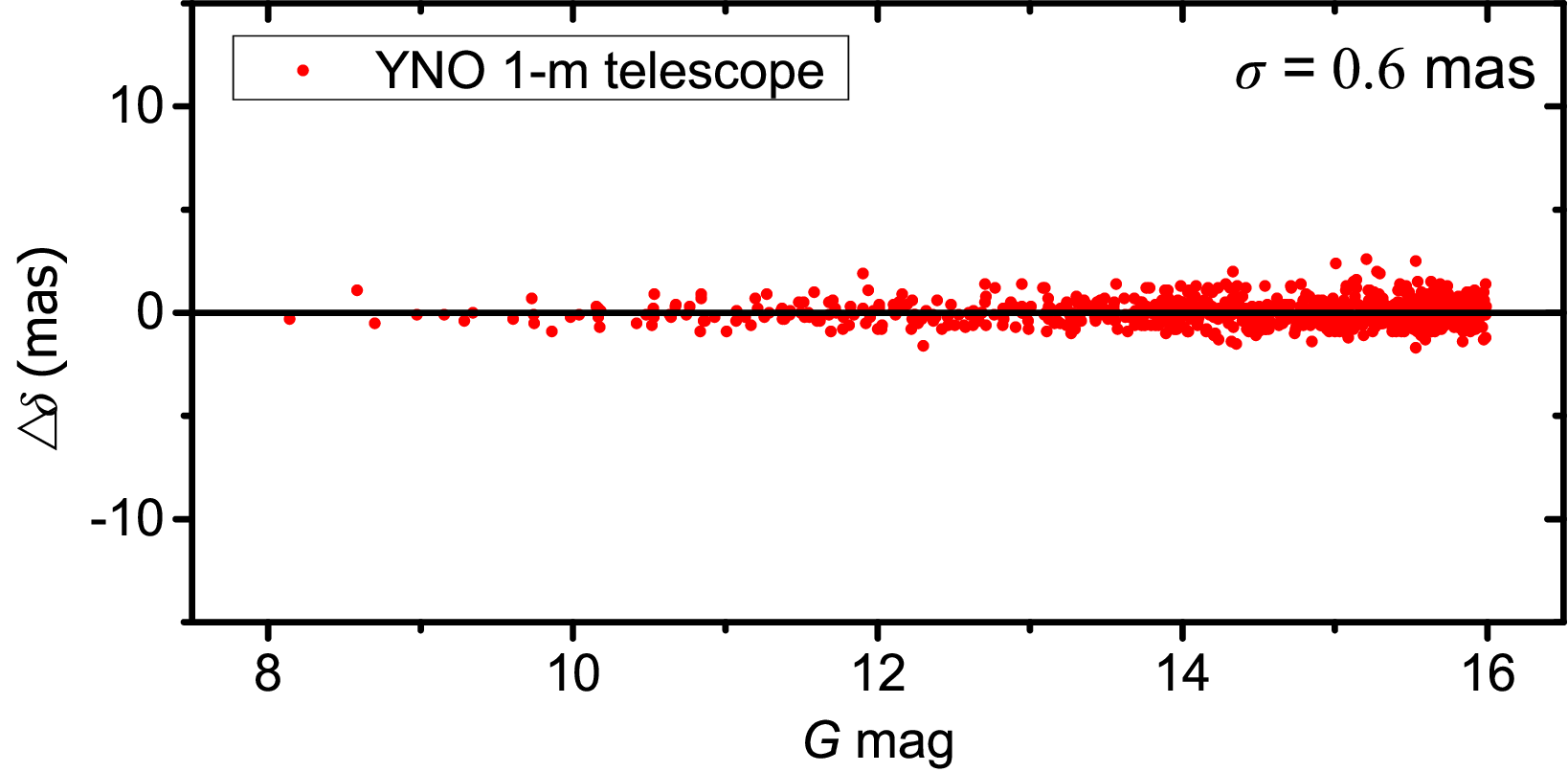}  
	\includegraphics[width=0.47   \columnwidth]{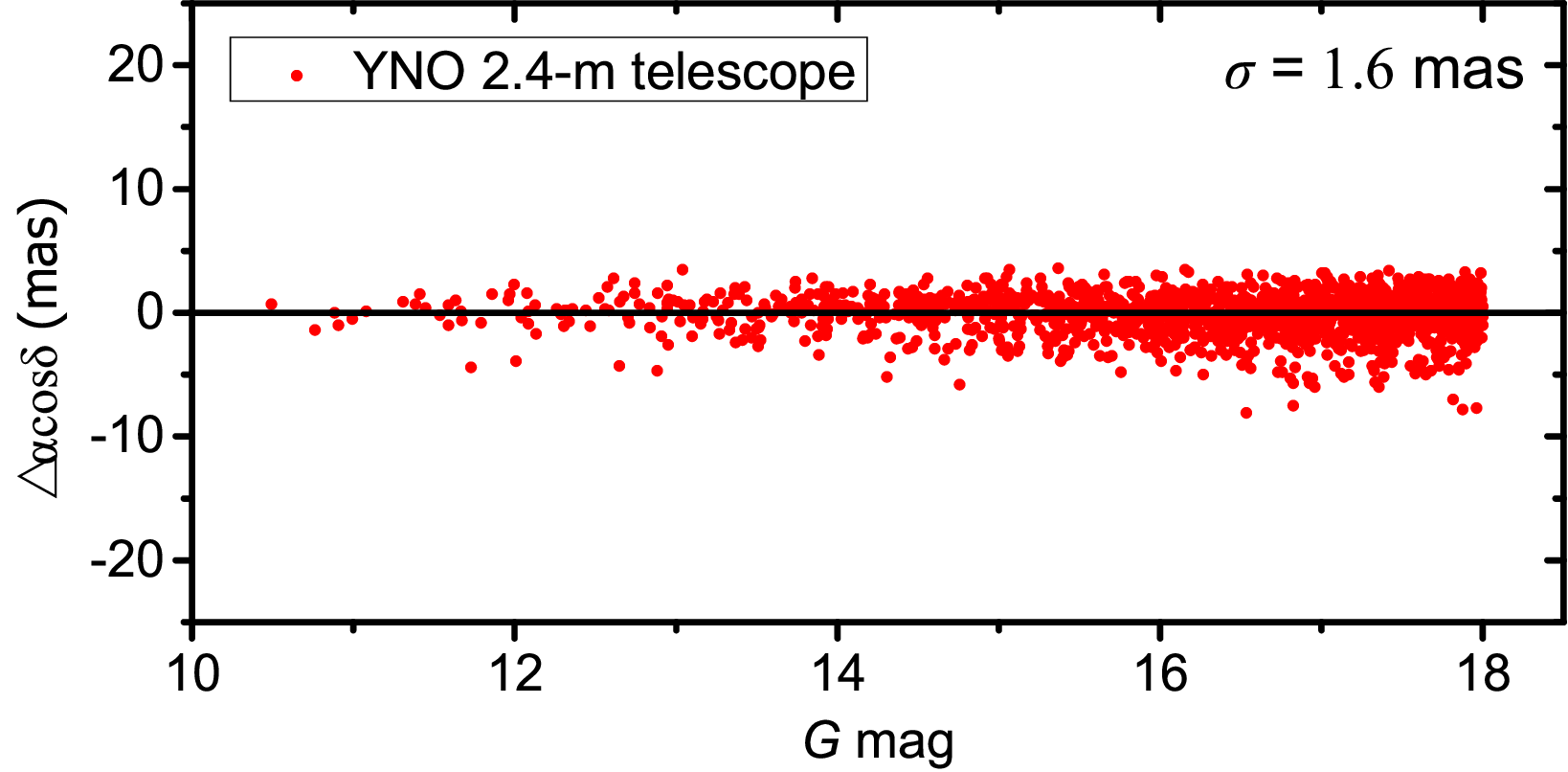} 
	\includegraphics[width=0.47  \columnwidth]{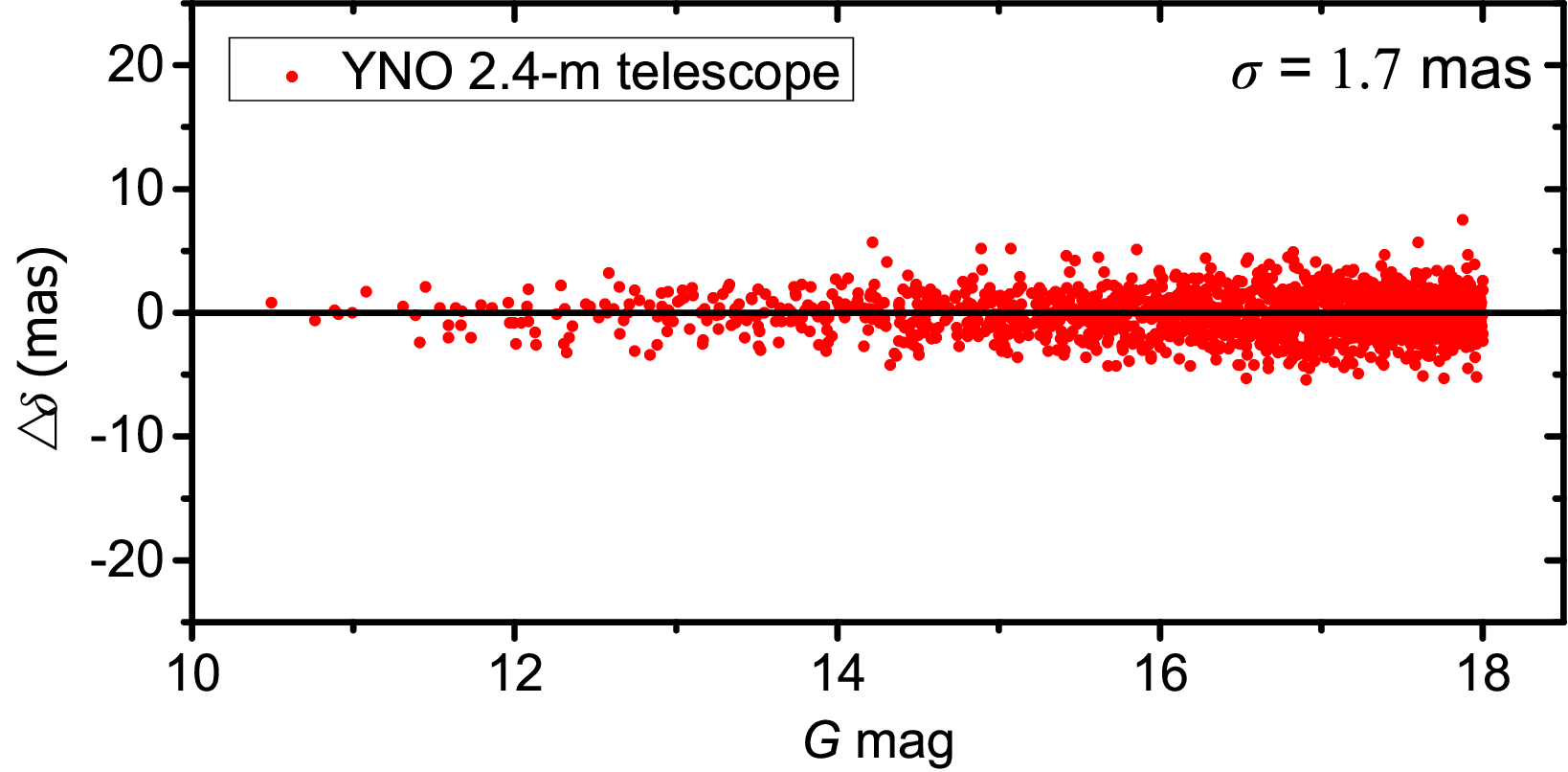}  
    \caption{The differences of the mean $(O-C)$ in R.A. and Dec. directions between our results and the reference results. The upper panels are the results for observation set 1 and the bottom panels are for observation set 2. Stars with low signal-to-noise ratios are excluded. The standard deviation of these differences $\sigma$ is shown in each panel.}
    \label{fig:accuracy}
\end{center}
\end{figure*}

After GD correction, the six-parameter plate constants are used for the reduction of observations to obtain astrometric results. The astrometric results corrected by the method of \citet{Wang2019} are used as a reference hereafter. Figure \ref{fig:precision} shows the positional standard deviation (SD) of each star, which is calculated as $\sigma_{\text{sum}}=\scriptstyle{\sqrt{\sigma_{\alpha \cos \delta}^2+\sigma_\delta^2}}$. In addition, the difference in the mean $(O-C)$ (i.e., the residual between the observed and computed position) between our results and the reference results is shown in Figure \ref{fig:accuracy}. As can be seen from Figure \ref{fig:precision} and \ref{fig:accuracy}, the astrometric results corrected using our GD solution are consistent with the reference results. On the one hand, the SDs of the astrometric results obtained by the two methods are equivalent. On the other hand, for the YNO 1-m telescope observations, which is less affected by GD, the mean $(O-C)$ difference between our results and the reference results is merely 1 mas. The difference is only 2 mas for the observations captured by the YNO 2.4-m telescope. That is to say, the method proposed in this paper can efficiently correct GD and obtain reliable astrometric results.

\subsection{Application of the GD solution}

As stated in Section \ref{intro}, our method is particularly useful in scenarios where only a limited number of bright reference stars (typically a dozen or so) can be used in the reduction. This usually happens when observing a sparse FOV, where the moving target may pass through. In this section, we processed and analyzed the observations of two targets that satisfy the scenario. These observations were not taken with a dithered FOV. Hence, the aforementioned well-established GD solutions are not applicable. 

\begin{figure} 
  
	 \includegraphics[width= 0.49  \columnwidth]{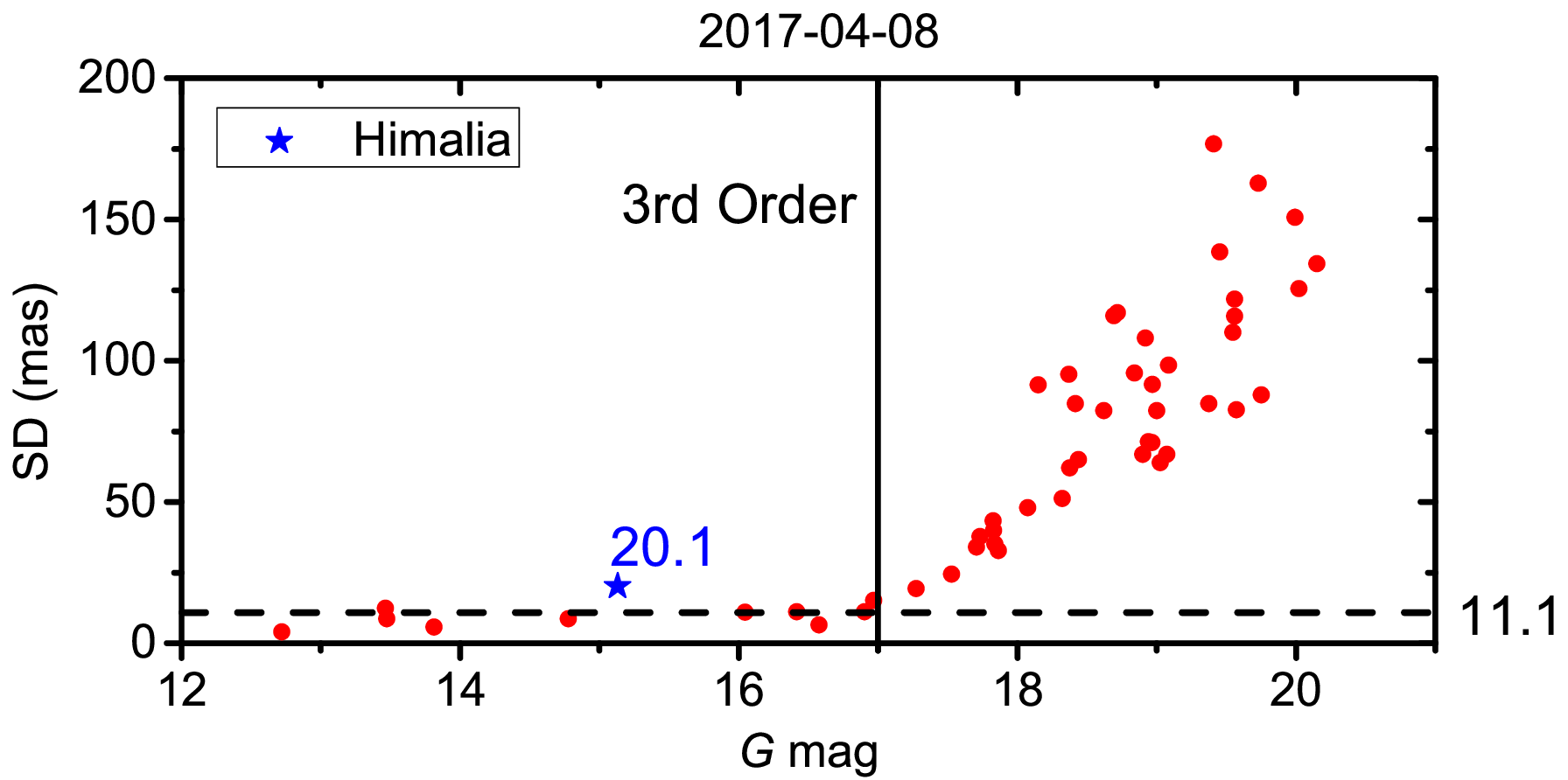}  
	\includegraphics[width= 0.49   \columnwidth]{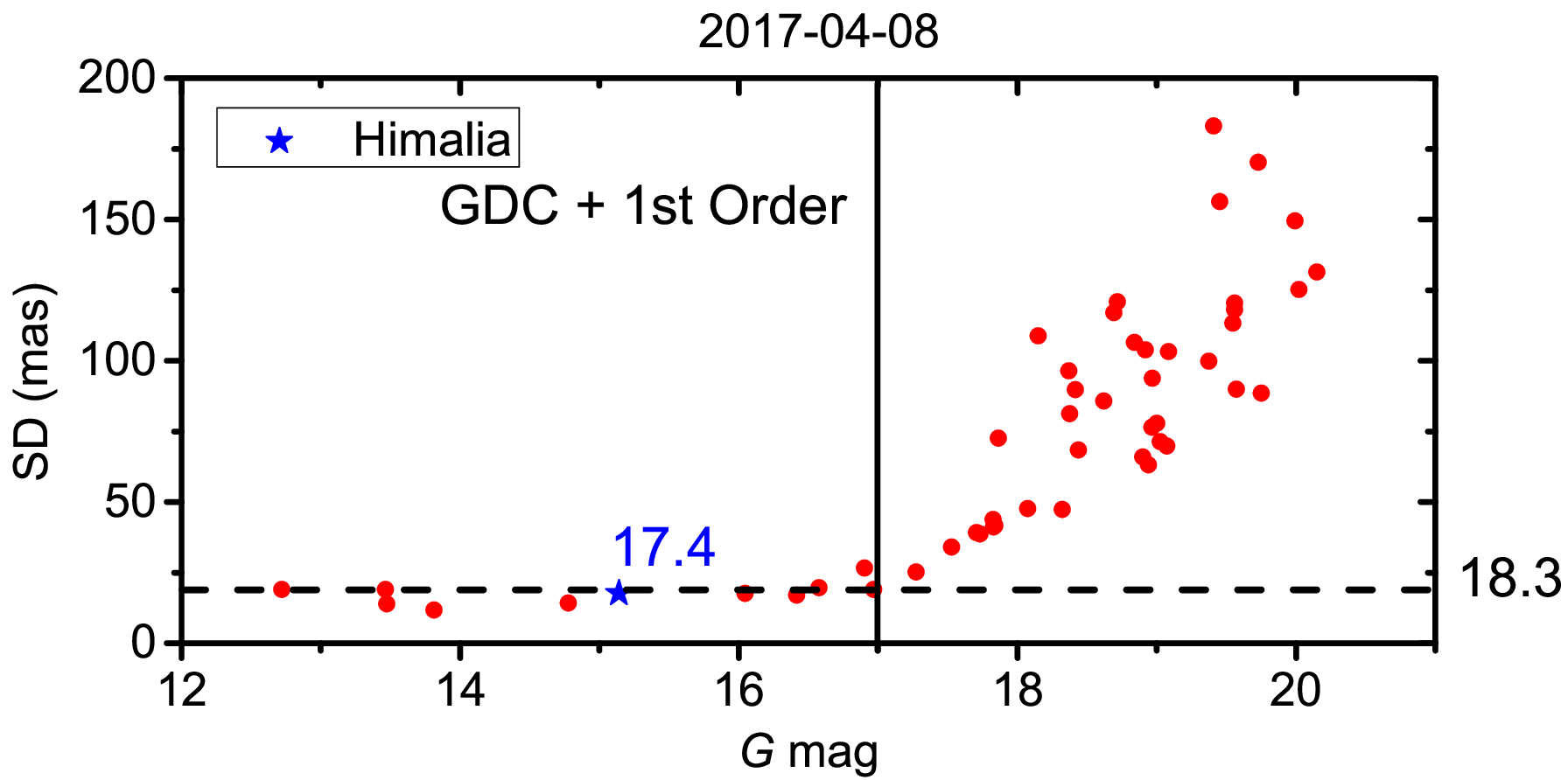}  
    \caption{The positional SD of Himalia (J6) before and after the GD correction. Overfitting occurs when 3rd-order plate constants were used, so that the SD of the target (J6) is obviously greater than the reference stars in the left panel.}
    \label{fig:j6} 
\end{figure}

Figure~\ref{fig:j6} shows the astrometric results of the J6 observations captured by the YNO 2.4-m telescope, the left panel shows an obviously greater SD for the target than other bright stars. This is because there are insufficient reference stars available for reduction, leading to overfitting of the 3rd-order plate constants. The higher precision for the reference stars in the left panel is an illusion, as overfitting absorbs the residuals in the reduction. More severe overfitting will lead to poorer calibration, resulting in lower precision of the target.

To address this issue, we corrected the GD corresponding to the 3rd-order polynomial using the method proposed in this work, and then used the six-parameter plate constants for reduction. The astrometric precision of the target J6 is improved after GD correction, with the positional SD decreased from 20 mas to 17 mas. The result is shown in the right panel of Figure~\ref{fig:j6}. 

\begin{figure} 
	\center
	\includegraphics[width=  0.44 \columnwidth]{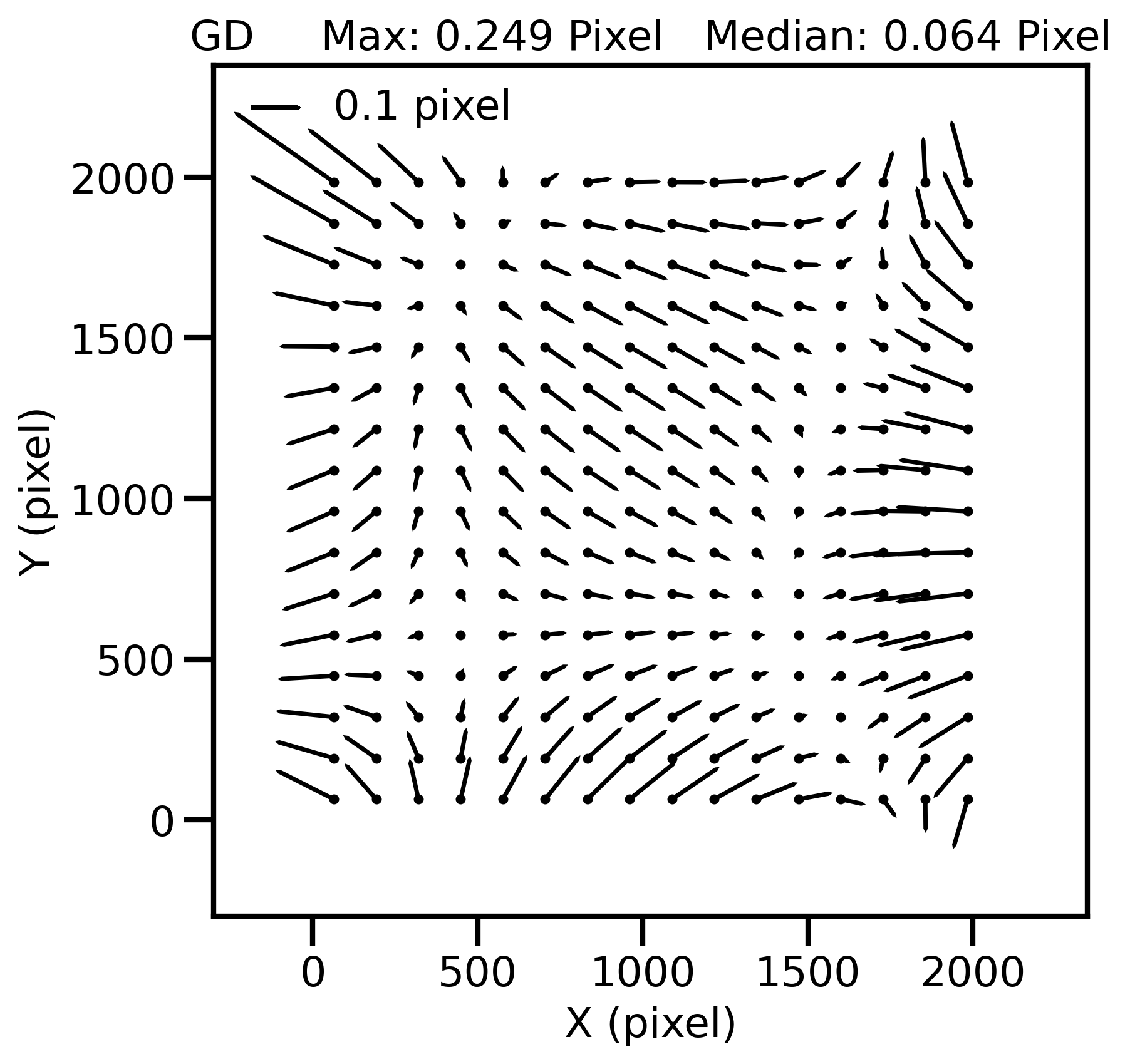}  \\ 
    \caption{The GD model of YNO 60-cm telescope.}
    \label{fig:gd60cm} 
\end{figure}

\begin{figure*}
\begin{center}
	\includegraphics[width= 0.48  \columnwidth]{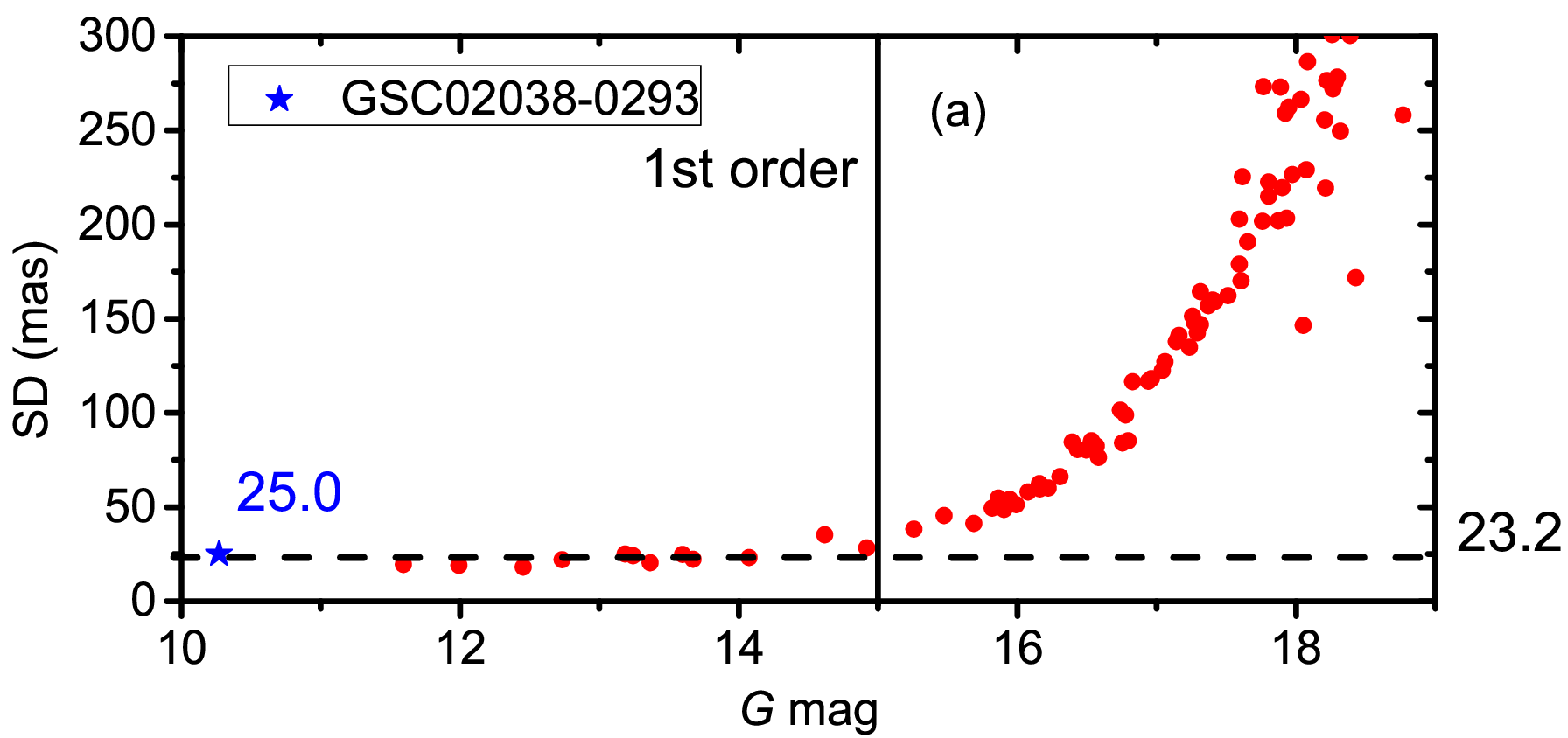}  
	\includegraphics[width= 0.48  \columnwidth]{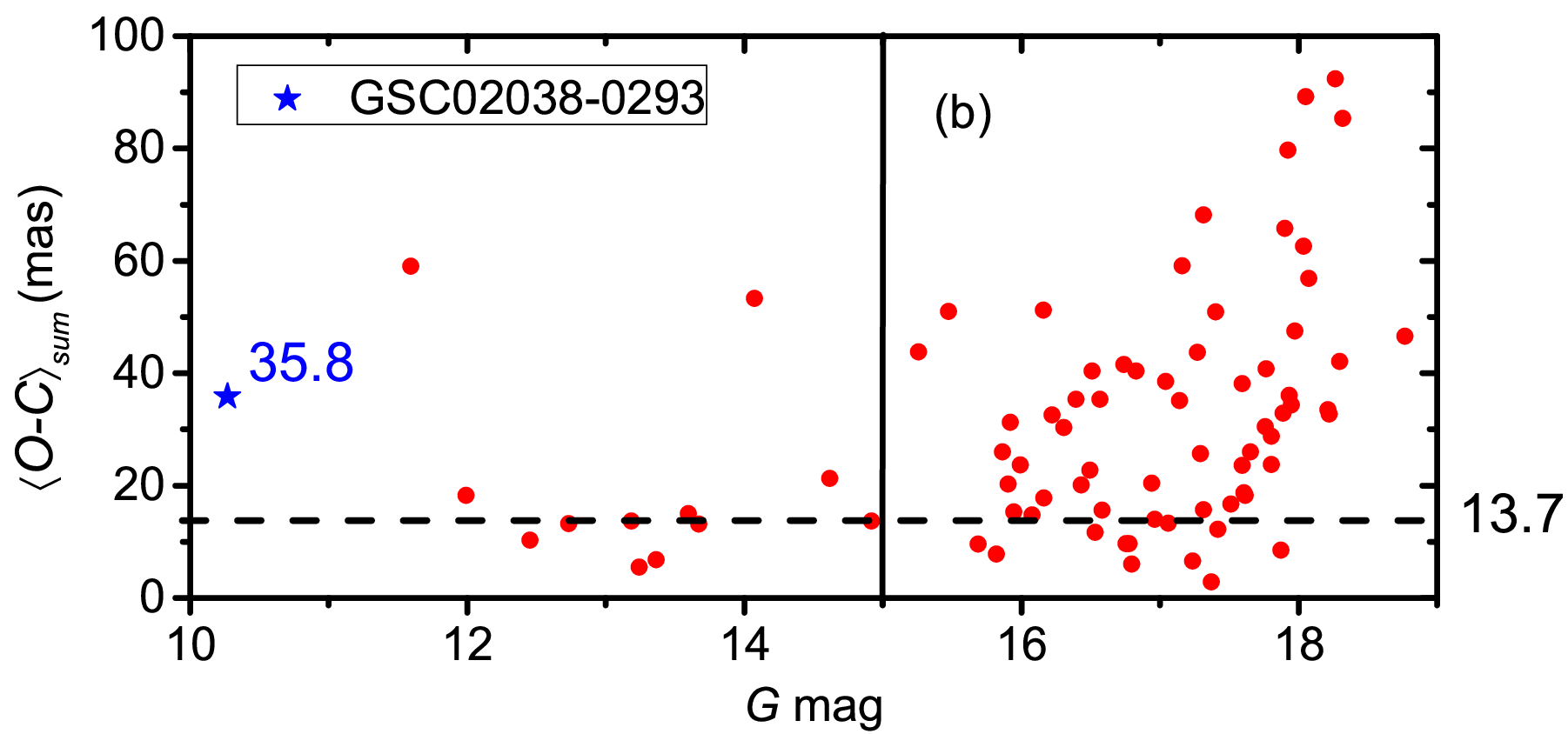}  
	\includegraphics[width= 0.48  \columnwidth]{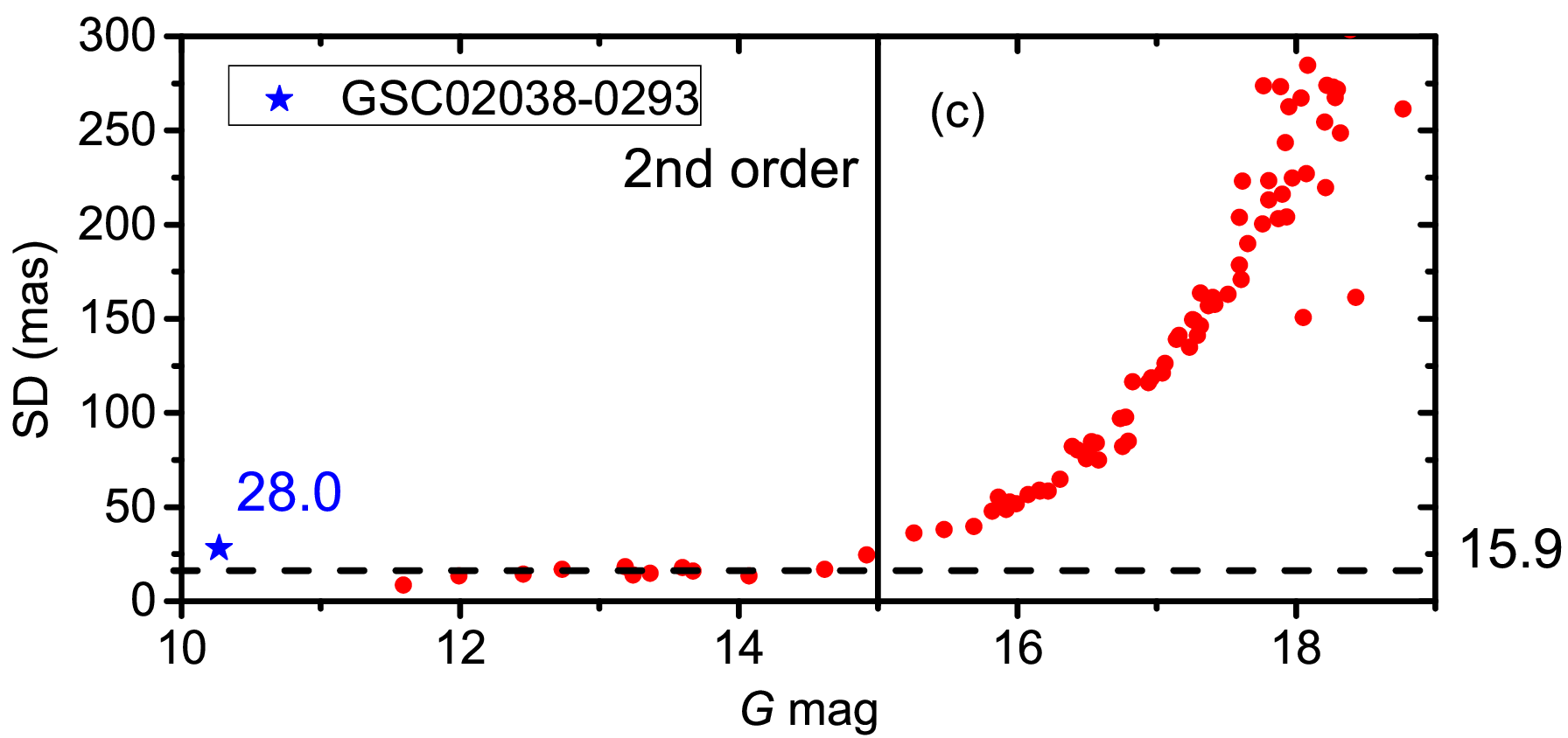}  
	\includegraphics[width= 0.48  \columnwidth]{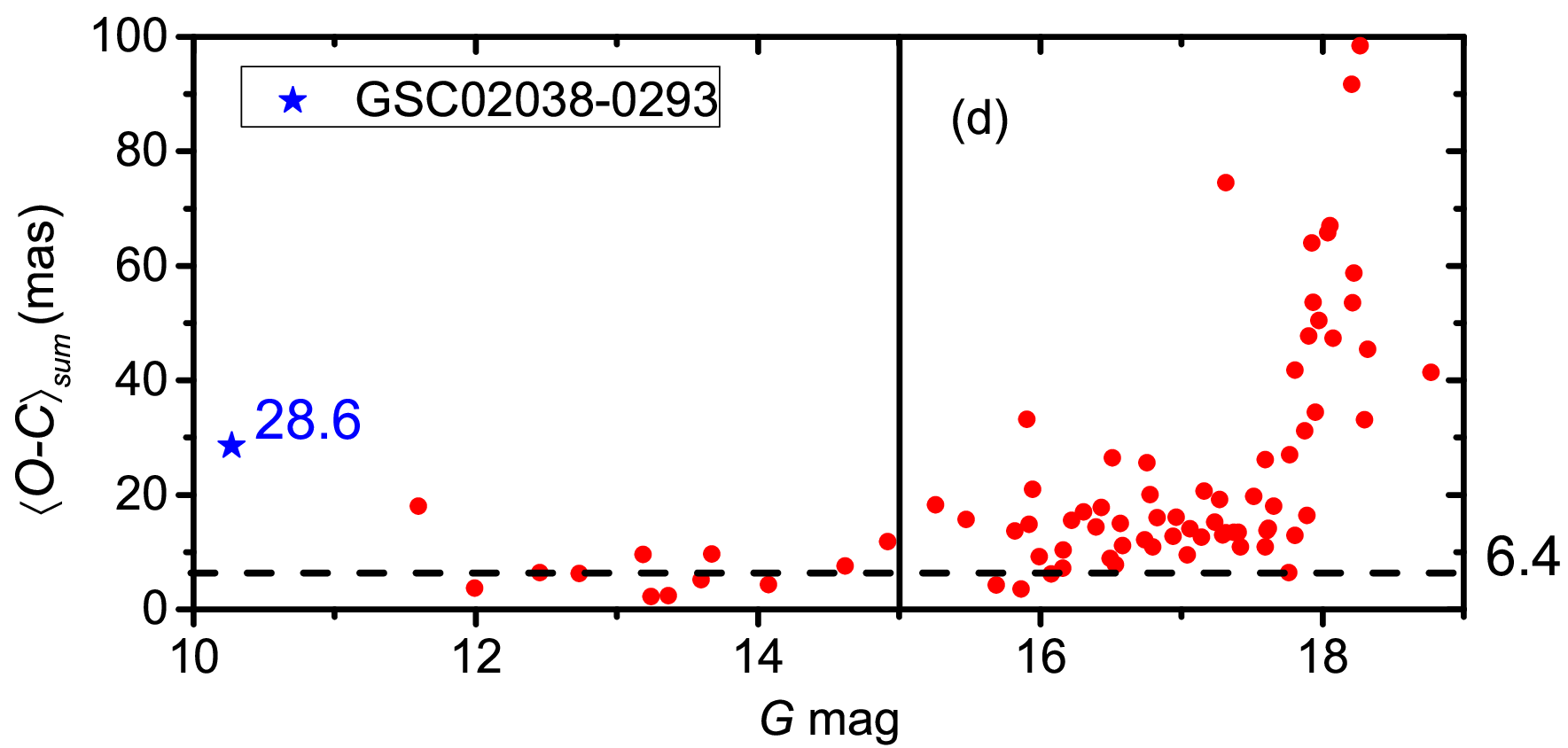} 
	\includegraphics[width= 0.48 \columnwidth]{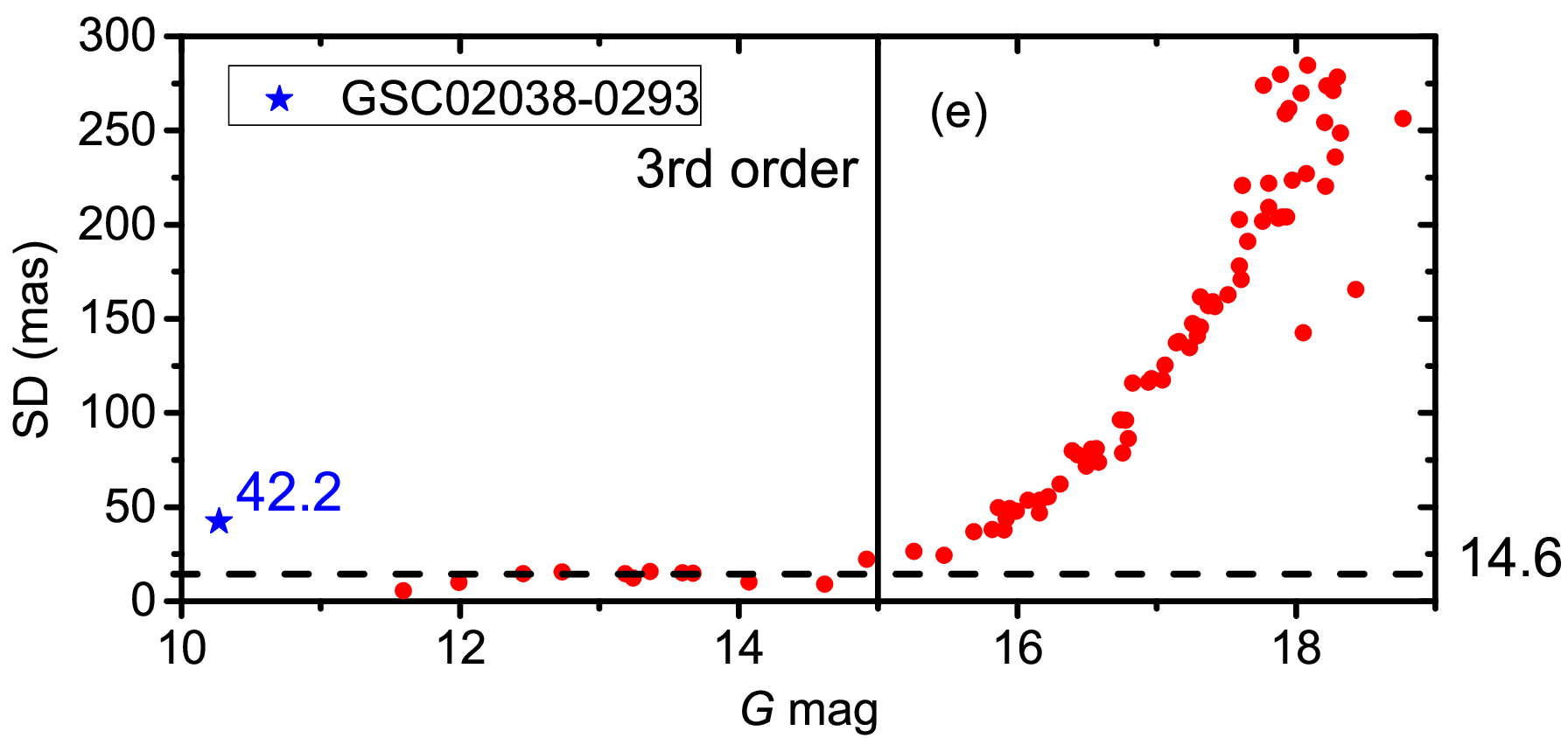}  
	\includegraphics[width= 0.48  \columnwidth]{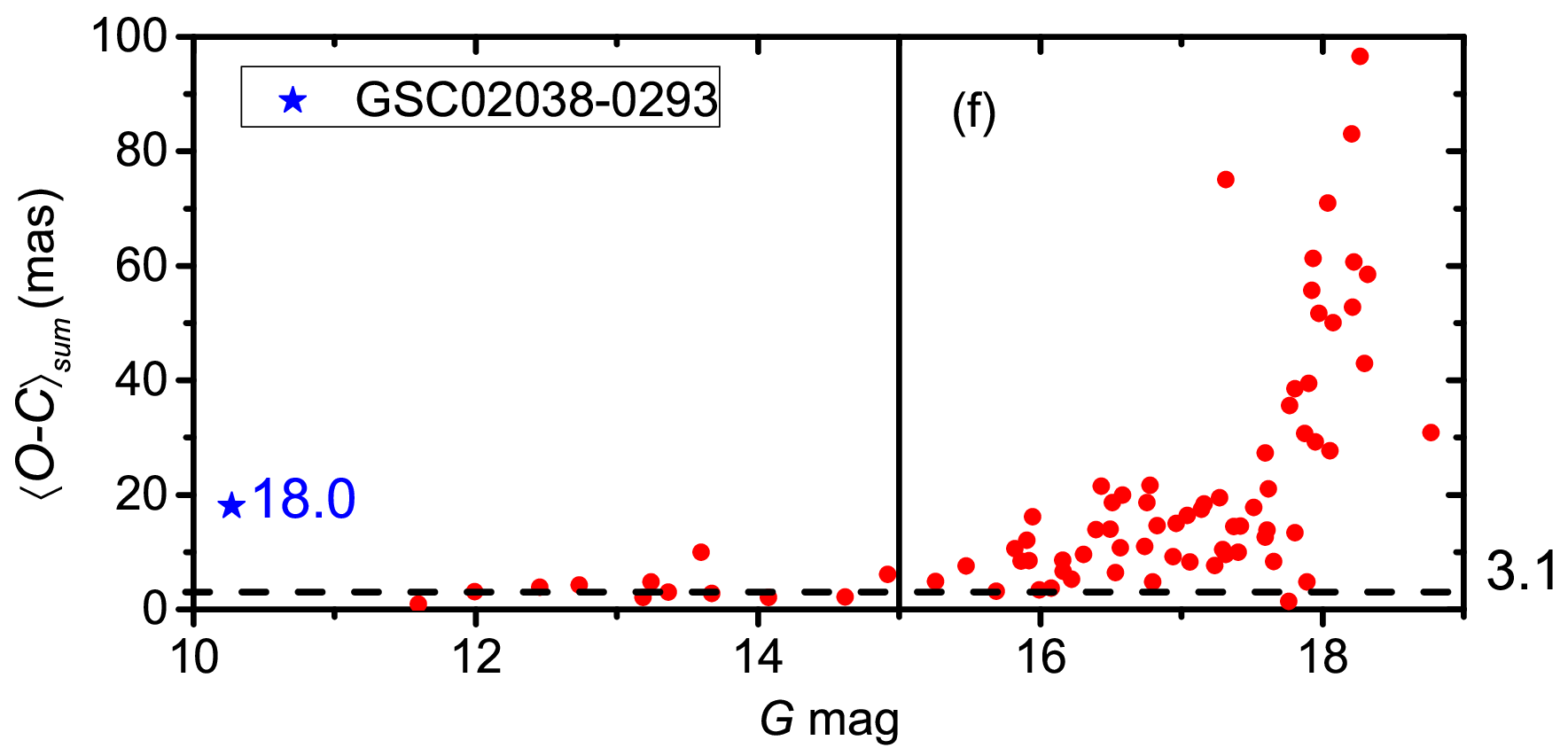} 
	\includegraphics[width= 0.48  \columnwidth]{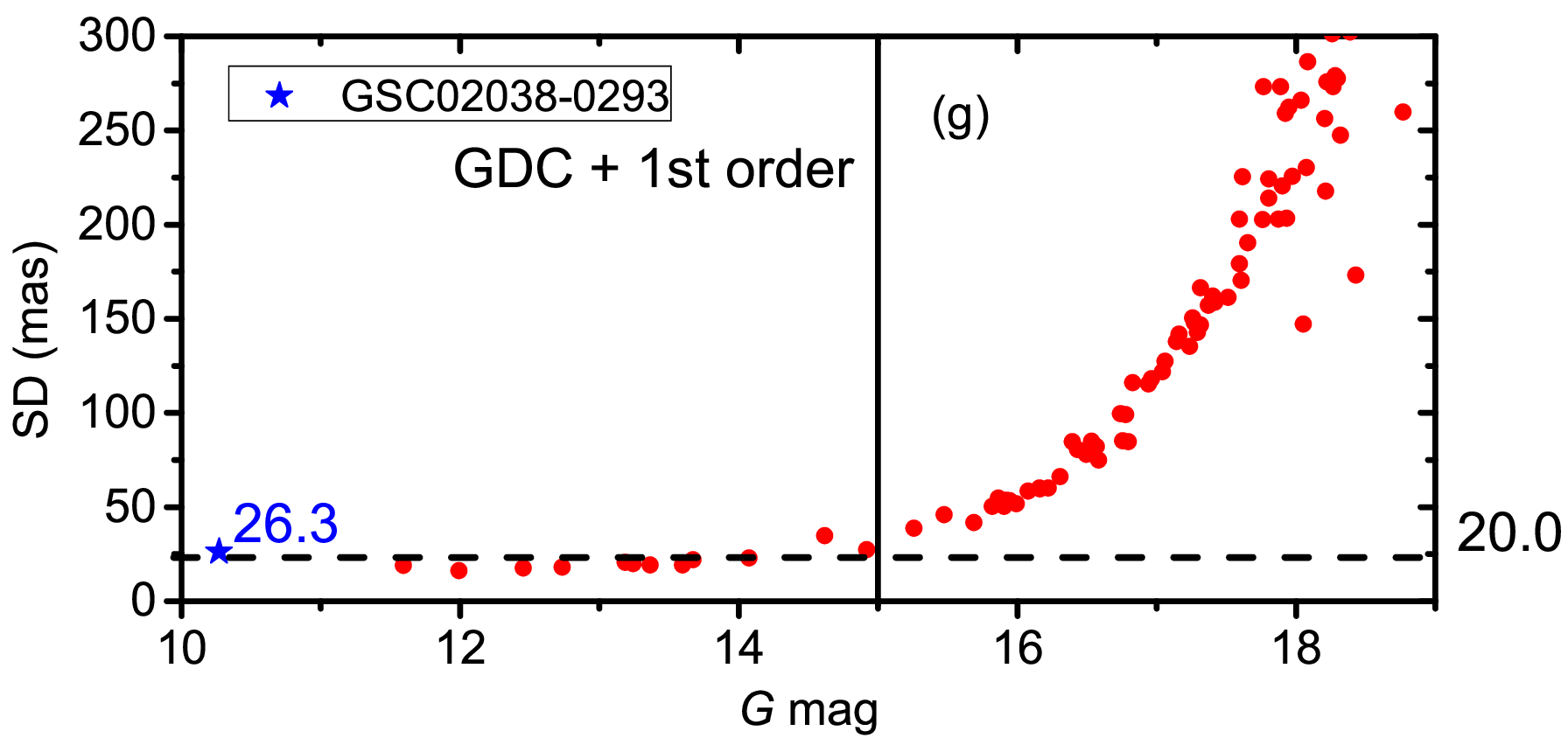}  
	\includegraphics[width= 0.48 \columnwidth]{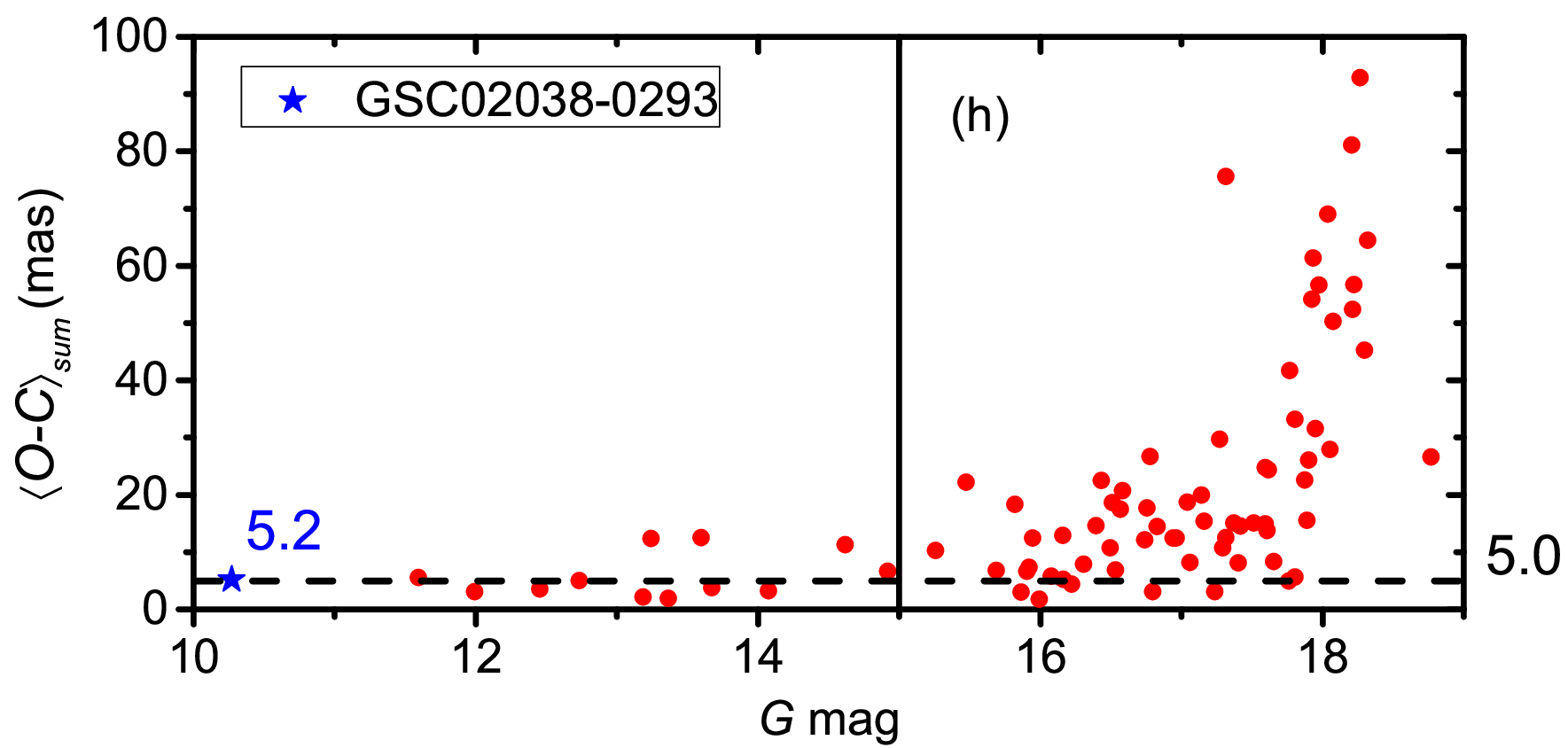}  

    \caption{Comparison of the positional SD and mean $(O-C)$ obtained by using different methods for reduction. From top to bottom in this figure are the results of reduction using 1st, 2nd and 3rd-order plate constants, as well as the results obtained by reduction using the six-parameter plate constants after GD correction. The left panels show the positional SD for each star, while the right panels present the corresponding mean $(O-C)$. The numbers in the panels are the statistics of the target GSC02038-00293. The horizontal dashed line marks the median of the positional SD or mean $(O-C)$ in each panel for stars brighter than 15 magnitude. }
    \label{fig:gsc2038}
\end{center}
\end{figure*}

The improvement is more significant for the observations taken with the YNO 60-cm telescope. Figure~\ref{fig:gd60cm} shows the GD model of the telescope solved with observation set 4. The GD model is characterized by a 3rd-order polynomial. To obtain reliable astrometric results, this GD solution was applied in the reduction of observation set 4. For comparison, Figure~\ref{fig:gsc2038} also presents the results obtained by using plate constants of different order for reduction. The left panel in the figure gives the positional SD and the right panel the corresponding mean $(O-C)$ calculated by $\langle O-C\rangle_{\mathrm{sum}}=\scriptstyle{\sqrt{\langle O-C\rangle_{\alpha \cos \delta}^2+\langle O-C\rangle_\delta^2}}$. 

Due to the observation set points to a fixed FOV, using low-order plate constants for reduction is possible to achieve precise positional measurement for the target (see 25.0 mas in panel (a) of Figure~\ref{fig:gsc2038}). However, despite the good fit of the plate constants at this time, the results for all stars displayed in the right panel (b) show large mean $(O-C)$, suggesting the presence of significant GD effect and rendering these astrometric results unreliable. As the order of the plate constants increased, the positional SD of the target given in panels (c) and (e) becomes greater than that of other bright stars. That is to say, even if only 2nd-order plate constants are used, overfitting will occur and become more pronounced as the order increases. This is consistent with the previous astrometric results of J6 observations. Additionally, panels (d) and (f) show that the mean $(O-C)$ values of the reference stars are decreased. This is due to overfitting resulting in the residual being absorbed in the reduction process. The target should not be involved in the fitting of the plate constants, so its mean $(O-C)$ values remain large, indicating unsatisfactory astrometric results.

The bottom two panels in Figure~\ref{fig:gsc2038} provide the results of applying our GD correction method first, followed by reduction using the six-parameter plate constants. As evident from the panels, these results show significant advantages compared to the results of other methods. On the one hand, the astrometric precision of the target is comparable to that achieved using the low-order plate constants. On the other hand, panel (h) reveals that the mean $(O-C)$ values for the target and reference stars are significantly smaller than the values obtained using the 1st and 2nd-order plate constants. This demonstrates that the system error caused by GD is significantly decreased after applying the GD correction.

It can be seen that for observations lacking sufficient reference stars to solve high-order plate constants, the GD solution significantly improves the astrometric results.
Empirically, when using the weighted least squares method to determine the plate constants for data reduction, the number of bright reference stars should be approximately 1.5 times the number of fitting parameters. This is crucial when determining the necessity of a GD solution.

\section{Conclusions and discussions}

A GD correction method based on the high-precision Gaia catalogue is investigated in this work. This method is effective and easy to implement. We presented the reduction of open clusters observations taken with the 1-m and 2.4-m telescopes at Yunnan Observatory to evaluate its accuracy. In the reduction, our method and a well-established method \citep{Peng2012, Wang2019} were used for GD correction, respectively. The results demonstrate that both methods achieve the same precision. In addition, the mean $(O-C)$ difference between our results and the reference results is only 1 mas for the YNO 1-m telescope observations and 2 mas for the YNO 2.4-m telescope observations. More observations were reduced to investigate the conditions necessary for our method. It was found that no more than fifteen frames are enough to derive the effective GD solution. These frames can have different or identical FOV, as long as each contains a sufficient number of reference stars that are approximately evenly distributed in the image. Typically, it is sufficient for the number of bright reference stars (with SNR$\ge$100) to be more than half the number of GD model parameters. As fitting errors are eliminated by averaging the coefficients from multiple frames, the final GD solution does not have overfitting issues even if only a dozen bright stars can be used to solve the high-order polynomial.

The major advantage of this method is it does not require special calibration observations to solve GD. This is of significant value for the historical observations where GD correction is unattainable due to the absence of relevant calibration data. Limited by the performance of observation equipment, even using the Gaia DR3 catalog, there are still not enough reference stars in these historical observations to solve high-order plate constants. Furthermore, reduction using 1st-order plate constants will result in significant systematic errors due to the severe effects of GD. The observations of the binary GSC02038-00293 taken with the YNO 60-cm telescope are an example that satisfies the situation. By applying our GD solution in the reduction, the astrometric results of this target were significantly improved, with the mean $(O-C)$ decreased from 36 mas to 5 mas. Additionally, the J6 observations in a sparse FOV taken with the YNO 2.4-m telescope were also corrected by the method. These results show that the method has great potential in improving the astrometric precision of the target in historical observations. 

It should be pointed out that we use a simpler and more easily calculated topocentric astrometric position of a reference star to drive GD, and found that there was no precision loss for our observations. Nevertheless, when establishing the GD model, it is more appropriate to adopt the observed positions of the reference stars, which can eliminate the influence from the differential atmospheric refraction and light aberration. The observed positions will be adopted in our future work to obtain accurate GD solutions.

\section*{Acknowledgements}

This work was supported by the National Key R\&D Program of China (Grant No. 2022YFE0116800), by the National Natural Science Foundation of China (Grant Nos. 12203019), by the Natural Science Foundation of Jiangxi Province (Grant Nos. 20242BAB20033), by the National Natural Science Foundation of China (Grant Nos. 11873026, 11273014) by the China Manned Space Project (Grant No. CMS-CSST-2021-B08) and the Joint Research Fund in Astronomy (Grant No. U1431227). The authors would like to thank the chief scientist Qian S. B. of the 1-m telescope and his working group for their kindly support and help. And thank them for sharing the observations of the binary GSC02038-00293. This work has made use of data from the European Space Agency (ESA) mission \emph{Gaia} (\url{https://www.cosmos.esa.int/gaia}), processed by the \emph{Gaia} Data Processing and Analysis Consortium (DPAC, \url{https://www.cosmos.esa.int/web/gaia/dpac/consortium}). Funding for the DPAC has been provided by national institutions, in particular the institutions participating in the \emph{Gaia} Multilateral Agreement.

\bibliographystyle{raa}
\bibliography{gdc}

\begin{thebibliography}{28}
\providecommand\natexlab[1]{#1}
\providecommand\JournalTitle[1]{#1}

\bibitem[{Anderson} {et~al.}(2006)]{Anderson2006}
{Anderson}, J., {Bedin}, L.~R., {Piotto}, G., {Yadav}, R.~S., \& {Bellini}, A.
  2006, Astronomy and Astrophysics, 454, 1029

\bibitem[{Anderson} \& {King}(2003)]{Anderson2003}
{Anderson}, J., \& {King}, I.~R. 2003, Publications Of The Astronomical Society
  Of The Pacific, 115, 113

\bibitem[{Bellini} \& {Bedin}(2009)]{Bellini2009}
{Bellini}, A., \& {Bedin}, L.~R. 2009, Publications Of The Astronomical Society
  Of The Pacific, 121, 1419

\bibitem[{Bellini} \& {Bedin}(2010)]{Bellini2010}
{Bellini}, A., \& {Bedin}, L.~R. 2010, Astronomy and Astrophysics, 517, A34

\bibitem[{Bernard} {et~al.}(2018)]{Bernard2018}
{Bernard}, A., {Neichel}, B., {Mugnier}, L.~M., \& {Fusco}, T. 2018, Monthly
  Notices of the Royal Astronomical Society, 473, 2590

\bibitem[{Casetti-Dinescu} {et~al.}(2021)]{Casetti2021}
{Casetti-Dinescu}, D.~I., {Girard}, T.~M., {Kozhurina-Platais}, V., {et~al.}
  2021, Publications Of The Astronomical Society Of The Pacific, 133, 064505

\bibitem[{Dal} {et~al.}(2012)]{Dal2012}
{Dal}, H.~A., {Sipahi}, E., \& {{\"O}zdarcan}, O. 2012, Publications of the
  Astronomical Society of Australia, 29, 150

\bibitem[{Gaia Collaboration} {et~al.}(2023)]{Gaia2023}
{Gaia Collaboration}, {Vallenari}, A., {Brown}, A.~G.~A., {et~al.} 2023,
  Astronomy and Astrophysics, 674, A1

\bibitem[{Grav} {et~al.}(2015)]{Grav2015}
{Grav}, T., {Bauer}, J.~M., {Mainzer}, A.~K., {et~al.} 2015, The Astrophysical
  Journal, 809, 3

\bibitem[{Green}(1985)]{Green1985}
{Green}, R.~M. 1985, {Spherical Astronomy}

\bibitem[{Illingworth} {et~al.}(2013)]{Illingworth2013}
{Illingworth}, G.~D., {Magee}, D., {Oesch}, P.~A., {et~al.} 2013, The
  Astrophysical Journal Supplement, 209, 6

\bibitem[{Kaplan} {et~al.}(1989)]{Kaplan1989}
{Kaplan}, G.~H., {Hughes}, J.~A., {Seidelmann}, P.~K., {Smith}, C.~A., \&
  {Yallop}, B.~D. 1989, The Astronomical Journal, 97, 1197

\bibitem[{Lin} {et~al.}(2019)]{Lin2019}
{Lin}, F.~R., {Peng}, J.~H., {Zheng}, Z.~J., \& {Peng}, Q.~Y. 2019, Monthly
  Notices of the Royal Astronomical Society, 490, 4382

\bibitem[{Lin} {et~al.}(2020)]{Lin2020}
{Lin}, F.~R., {Peng}, Q.~Y., \& {Zheng}, Z.~J. 2020, Monthly Notices of the
  Royal Astronomical Society, 498, 258

\bibitem[{McKay} \& {Kozhurina-Platais}(2018)]{McKay2018}
{McKay}, M., \& {Kozhurina-Platais}, V. 2018, {WFC3/IR: Time Dependency of
  Linear Geometric Distortion}, Instrument Science Report WFC3 2018-9, 11 pages

\bibitem[{Peng} {et~al.}(2017)]{Peng2017}
{Peng}, H.~W., {Peng}, Q.~Y., \& {Wang}, N. 2017, Monthly Notices of the Royal
  Astronomical Society, 467, 2266

\bibitem[{Peng} \& {Fan}(2010)]{Peng2010}
{Peng}, Q.~Y., \& {Fan}, L.~Y. 2010, Chinese Science Bulletin, 55, 791

\bibitem[{Peng} {et~al.}(2012)]{Peng2012}
{Peng}, Q.~Y., {Vienne}, A., {Zhang}, Q.~F., {et~al.} 2012, The Astronomical
  Journal, 144, 170

\bibitem[Peng {et~al.}(2015)]{Peng2015}
Peng, Q.~Y., Wang, N., Vienne, A., {et~al.} 2015, Monthly Notices of the Royal
  Astronomical Society, 449, 2638

\bibitem[{Peng} {et~al.}(2023)]{Peng2023}
{Peng}, X.~Y., {Qi}, Z.~X., {Zhang}, T.~M., {et~al.} 2023, The Astronomical
  Journal, 165, 172

\bibitem[{Reid} \& {Menten}(2007)]{Reid2007}
{Reid}, M.~J., \& {Menten}, K.~M. 2007, The Astrophysical Journal, 671, 2068

\bibitem[{Service} {et~al.}(2016)]{Service2016}
{Service}, M., {Lu}, J.~R., {Campbell}, R., {et~al.} 2016, Publications Of The
  Astronomical Society Of The Pacific, 128, 095004

\bibitem[{Shang} {et~al.}(2022)]{Shang2022}
{Shang}, Y.~J., {Peng}, Q.~Y., {Zheng}, Z.~J., {et~al.} 2022, The Astronomical
  Journal, 163, 210

\bibitem[{Wang} {et~al.}(2017)]{Wang2017}
{Wang}, N., {Peng}, Q.~Y., {Peng}, H.~W., {et~al.} 2017, Monthly Notices of the
  Royal Astronomical Society, 468, 1415

\bibitem[{Wang} {et~al.}(2019)]{Wang2019}
{Wang}, N., {Peng}, Q.~Y., {Zhou}, X., {Peng}, X.~Y., \& {Peng}, H.~W. 2019,
  Monthly Notices of the Royal Astronomical Society, 485, 1626

\bibitem[{Zang} {et~al.}(2022)]{Zang2022}
{Zang}, L., {Qian}, S.~B., {Zhu}, L.~Y., \& {Liu}, L. 2022, Monthly Notices of
  the Royal Astronomical Society, 511, 553

\bibitem[{Zhai} {et~al.}(2018)]{Zhai2018}
{Zhai}, C.~X., {Shao}, M., {Saini}, N.~S., {et~al.} 2018, The Astronomical
  Journal, 156, 65

\bibitem[{Zheng} {et~al.}(2021)]{Zheng2021}
{Zheng}, Z.~J., {Peng}, Q.~Y., \& {Lin}, F.~R. 2021, Monthly Notices of the
  Royal Astronomical Society, 502, 6216

\end{thebibliography}

\end{document}